\documentclass[12pt]{article}
\usepackage[pdftex]{graphicx}
\usepackage{wrapfig}
\usepackage{latexsym}
\usepackage[pdftex]{color}
\usepackage{pifont}
\usepackage{epstopdf}
\usepackage{tikz}
\usetikzlibrary{decorations.pathmorphing,patterns}

\newcommand{\bmp}[2][t]{\begin{minipage}[#1]{#2}}
\newcommand{\emp}{\end{minipage}}

\usepackage{amsmath}
\usepackage{amssymb}
\usepackage[square,numbers,sort&compress]{natbib}

\begin{document}

\title{Energy output from a single outer hair cell}
\author{Kuni H. Iwasa\\
\normalsize{Department of Otolaryngology, Stanford University School of Medicine}\\
\normalsize{Stanford, California 94305}\\
\normalsize{and}\\
\normalsize{NIDCD, National Institutes of Health}\\
\normalsize{Bethesda, Maryland 20892}
}

\date{}

\maketitle

\section*{Abstract}
Electromotility of outer hair cells (OHCs) has been extensively studied with \emph{in vitro} experiments because of its physiological significance to the cochlear amplifier, which provides the exquisite sensitivity and frequency selectivity of the mammalian ear. However, these studies have been performed largely under load-free conditions or with static load, while these cells function \emph{in vivo} in a dynamic environment, receiving electrical energy to enhance mechanical oscillation in the inner ear.  This gap leaves uncertainties in addressing a key issue, how much mechanical energy an OHC provides. The present report is an attempt of bridging the gap by introducing a simple one-dimensional model for electromotility of OHC in a dynamic environment.  
This model incorporates a feedback loop involving the receptor potential and the mechanical load on OHC, and leads to an analytical expression for the membrane capacitance, which explicitly describes the dependence on the elastic load, viscous drag, and the mass. The derived equation of motion was examined in a mass-less model system with realistic parameter values for OHC. It was found that viscous drag is more effective than elastic load in enhancing the receptor potential that drives the cell. For this reason, it is expected that OHCs are more effective in counteracting viscous drag than providing elastic energy to the system.
\subsection*{key words:} amplifier, mammalian ear, membrane capacitance

\pagebreak

\section{Introduction}\label{sec:intro}
 Considerable progress has been made in recent years in our understanding of the mechanism of prestin-based somatic motility, or ``electromotility,'' of outer hair cells (OHCs) in the cochlea, both on cellular- \cite{a1987,s1991,Dong2013} and the molecular levels~\cite{zshlmd2000,oliv-fakl2001,Song2013}, as well as clarifying its physical basis that it is based on electromechanical coupling \cite{i1993,ts1995,i2001}. For example, experiments on isolated OHCs have determined load-free displacement~\cite{a1987,s1991} and isometric force production \cite{hal1995,ia1997}. These experimental observations can be described by static models \cite{dhe1993,i2001}. Nonetheless, these are the conditions under which those cells do not provide energy. Some theoretical works have addressed energy production by OHCs  \cite{Rabbitt2009,Ramamoorthy2012a} by extrapolating from these \emph{in vitro} conditions. However, these analyses do not provide an equation of motion or the dependence of nonlinear capacitance on external mechanical load, the essential features to describe the production of mechanical energy for amplifying acoustic signal.
 
There are a number of issues to be addressed for describing OHCs in a dynamic environment. One such issue is the frequency dependence of the motile response. The amplitude of displacement in response to voltage changes rolls off at about 15 kHz under load-free condition, while force production near isometric condition remains flat up to 60 kHz~\cite{fhg1999}. This difference likely indicates that the frequency response depends on the mechanical load, more specifically viscoelastic drag. 

Another issue is attenuation of the receptor potential by the membrane capacitance at operating frequencies~\cite{ha1992,Johnson2011,ospeck2012}. The membrane capacitance consists of two main components. One is structural and is proportional to the membrane area. Another is nonlinear component associated with the mobile charge of prestin, which flips in the electric field on conformational changes. Nonlinear capacitance has been described under load-free condition, at which it is expected to be the largest. Since constraint on the membrane area almost eliminates the nonlinear component~\cite{ai1999}, it could be expected that external load reduces this component and thus reduces capacitive current, which attenuates the receptor potential at higher frequencies.
This feedback would improve the performance of OHCs, particularly at high frequencies. This effect still needs to be described quantitatively. Previous treatments used either load-free capacitance~\cite{ha1992,odi2003a,Mistrik2009,maoil2013} or the linear capacitance alone~\cite{Johnson2011,ospeck2012}.

Here a model is presented for describing the motion of a single OHC under mechanical loads.   In the following, the basic equations are derived. That is followed by derivations of quantities that characterize the motile element. These quantities are determined using experimental values for OHC that operates at 4 kHz. Then the balance of energy input and output is examined for those cells under mass-free condition. The implications are discussed on those cells, which operate at higher frequencies.  

\section{The model}\label{sec:model}
\begin{figure}[t] 
\begin{center}
\includegraphics[height=6cm]{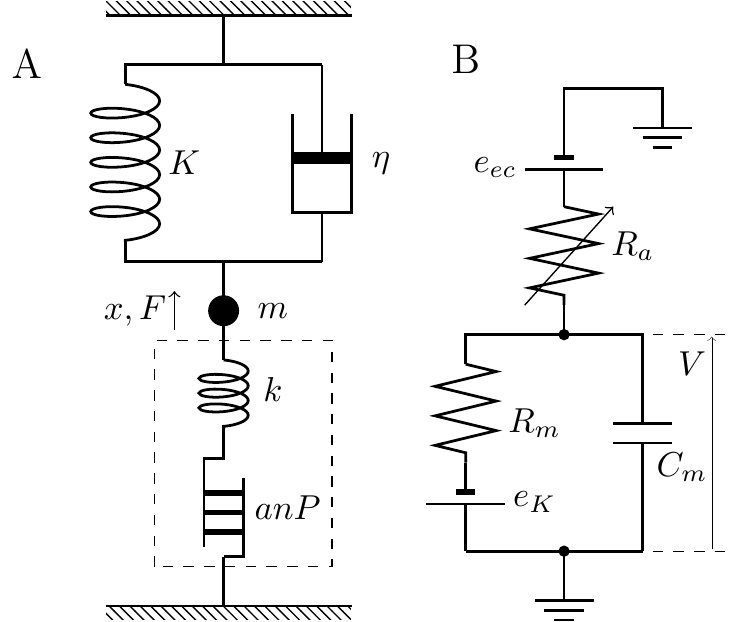}
\end{center}
\caption{\small{Mechanical connectivity and the equivalent electric circuit of the system examined. The system is driven by changes in hair bundle conductance $R_a$. Unlike \emph{in vivo} condition, movement of the cell body does not affect $R_a$. In the mechanical schematics (A), $K$ is stiffness of the external mechanical load, $m$ the mass, and positive force $F$ and positive displacement $x$ of the cell are upward. The drag coefficient is $\eta$. The contribution of the motile element to cell length is $anP$, where $P$, $a$, and $n$ respectively represent the fraction of the motile elements in the elongated state, unitary length change, and the number of such units, the unitary change of charge of which is $q$.  The stiffness of the cell due to the material property alone is $k$. The broken line indicates the border of the OHC. In the equivalent circuit (B) of the hair cell, the membrane potential is $V$, the hair bundle resistance $R_a$, the basolateral resistance $R_m$, and the total membrane capacitance of the basolateral membrane $C_m$, consisting of the structural capacitance $C_0$ and the contribution of charge movements in the motile element, which depends on the load (described in Section \ref{sec:cap}). The endocochlear potential is $e_{ec}$ and the potential $e_K$ is due to K$^+$ permeability of the basolateral membrane. The apical capacitance is ignored in this model. }}
\label{fig:system}
\end{figure}

Here we consider a simple system, which consists of an OHC, elastic load, drag, and a mass (Fig.\ \ref{fig:system}). This is not to approximate the organ of Corti, but to describe an OHC, which is subjected to mechanical load. This system with a single degree of freedom is described by writing down the equations for the motile mechanism, the receptor potential, and the equation of motion. These equations are interrelated and they constitute a set of simultaneous equations, which are examined in the subsequent sections.

\subsection{The motile mechanism}
Outer hair cells (OHCs) have a motile mechanism driven by the membrane potential based on mechanoelectric coupling. Here a one-dimensional model is used instead of a membrane model~\cite{i2001} for simplicity. We assume that the cell has $n$ motile elements, which has two discrete states, compact and extended, and during a transition from the compact state to the extended state, the cell length increases by $a$ and the electric charge $q$ flips across the plasma membrane. Let  $P$ be the fraction of the motile units in the extended state. Its equilibrium value $P_\infty$ follows the Boltzmann distribution,
\begin{eqnarray}\label{eq:distr}
P_\infty=\frac{\exp[-\beta\Delta G]}{1+\exp[-\beta\Delta G]},
\end{eqnarray}
with $\beta=1/(k_BT)$, where $k_B$ is Boltzmann's constant and $T$ the temperature, and 
\begin{equation}
\Delta G=q(V-V_{1/2})-aF
\label{eq:G}
\end{equation}
represents the difference in the free energy in the two states, referenced to the compact state. Here $F$ is  the force applied in the direction of extending the cell and $V$ the membrane potential. The quantity $V_{1/2}$ is a constant that determines the operating point. Here both $q$ and $a$ are positive because rising membrane potential and decreasing extensive force increases $\Delta G$, and thus reduces the fraction $P_\infty$ of the extended state.

Now consider a case, in which the motile mechanism is connected to an external elastic element (Fig.\ \ref{fig:system} without the mass or the dashpot). The force $F$ applied to the cell depends on the elastic elements as well as the conformational change of the motile element elicited to the voltage change. The elastic elements include the material stiffness $k$ of the cell  as well as an external elastic load $K$. Assume that at the membrane potential $V$ changed from its resting value $V_0$. The resulting displacement $x$ of the cell produces force $F$ on the external spring. The same force is applied to the cell reciprocally, producing a displacement $F/k (=-Kx/k)$. Thus the displacement $x$ is determined from $x=an(P-P_0)-Kx/k$, where the fraction $P$ of the extended state, is changed from its resting value $P_0$.
This change is expressed by,
\begin{eqnarray} \label{eq:x}
x&=&\tilde{K}/K\cdot an(P-P_0),\\ \label{eq:F}
F&=&-\tilde{K}an(P-P_0),
\end{eqnarray}
with effective spring constant $ \tilde{K}=k\cdot K/(k+K)$. Notice that an increase in $P$, which increases $x$, generates force in the contracting direction and that the maximal value of $\tilde{K}$ is $k$.
This force $F$ is applied to the motile element in the cell. This leads to the expression for the difference $\Delta G$ in the free energy of the two states, 
\begin{eqnarray}\label{eq:energy}
\Delta G=q(V-V_{1/2})+\tilde{K}a^2n(P-P_0).
\end{eqnarray}
Notice here that an increase in $P$ increases the energy $\Delta G$, making a further increase less favorable. If the system is in equilibrium, $P=P_\infty$ and the equilibrium value of $P_\infty$ is determined by  combining Eq.\ \ref{eq:distr} with Eq.\ \ref{eq:energy}, in which $P$ is substituted with $P_\infty$. Under this condition, all forces are balanced and no movement takes place.

Here we notice that $\Delta G$, and therefore $P_\infty$, can be determined for any set of $V$ and $P$ by combining Eqs.\ \ref{eq:distr} and \ref{eq:energy}. Since $P-P_\infty=0$ in equilibrium, the difference $P-P_\infty$ quantifies the deviation from equilibrium that moves the cell. 

\subsection{Equation of motion}
Now we examine how to describe the movement of the cell, which has mechanical loads (Fig.\ \ref{fig:system}).  The system has a single degree of freedom and is described by using length displacement $x$ of the cell as the variable. What is the force that drives the cell? 

Let the displacement be fixed at $x$ and the state of the motile element be $P_\infty$ in the beginning. Suppose the membrane potential suddenly changes at a certain moment. The cell generates force that moves the cell.  The force that drive the cell is $k\cdot an(P-P_\infty)$ if the stiffness of the cell is $k$ and if $P_\infty-P$ is small enough so that the linear term is dominant.  Even though the motile element contributes to reduce the stiffness from the intrinsic stiffness $k$, in Section \ref{sec:param} we confirm that this is indeed a good approximation for a set of parameter values that we use. Then, the equation of motion should be,
\begin{equation}
m\frac{d^2x}{dt^2}+\eta\frac{dx}{dt}=k\cdot an(P_\infty-P),
\label{eq:mx}
\end{equation} 
where $\eta$ is the drag coefficient, and $m$ the mass.

For a given displacement $x$, $P$ can be given by Eq.\ \ref{eq:x}. Since $P_0$ is a time-independent constant, its time derivatives does not contribute. The equation of motion can then be expressed,
\begin{eqnarray} \label{eq:motion}
m\frac{d^2P}{dt^2}+\eta\frac{dP}{dt}&=&(k+K)(P_\infty-P).
\end{eqnarray}
Notice that a factor $k\cdot na$ drops out from the equation because it is shared by all terms.  In the special case of $m=0$, Eq.\ \ref{eq:motion} turns into a relaxation equation with a time constant $\eta/(k+K)$, which is intuitive.

\subsection{Receptor potential}
The motile response of the cell is driven by the receptor potential, generated by the receptor current, which is, in turn, elicited by changes in the hair bundle resistance by mechanical stimulation.  This current is driven by the sum of two electromotive forces. One is  $e_K$, which is primarily determined by the K$^+$  conductance of the basolateral membrane of the cell. The other is the endocochlear potential $e_{ec}$, which is generated by the stria vascularis, a tissue that lines a part of the scala media  (Fig.\ \ref{fig:system}B). 

The magnitude of the receptor potential is determined not only by changes in the hair bundle conductance but also by the basolateral conductance because the electric current through the apical membrane must be equal to the current through the basolateral membrane. Changes in the membrane potential elicit not only ionic currents but also a capacitive current proportional to the regular capacitance in the basolateral membrane. In addition, they flip the charge of the motile units, produces an additional current, similar to the capacitive current. Thus, Kirchhoff's law leads to,
\begin{eqnarray} \label{eq:potential}
\frac{e_{ec}-V}{R_a}&=&\frac{V-e_K}{R_m}+C_0\frac{dV}{dt}-nq\frac{dP}{dt}.
\end{eqnarray}
Here $R_a$ is the apical membrane resistance, which is dominated by mechanotransducer channels in the hair bundle.   The basolateral membrane has the resistance $R_m$ and the linear capacitance $C_0$, which is determined by the membrane area and the specific membrane capacitance of $\sim 10^{-2}$ F/m$^2$ for biological membranes \cite{cole,sokabe-sachs1991}. The apical membrane capacitance is ignored for simplicity. 

The last term on the right-hand-side of Eq.\ \ref{eq:potential} is an additional displacement current due to the charge movement in the motile elements.  It has the negative sign because a voltage increase results in a decrease in $P$ as mentioned earlier.  As we will see later, nonlinear capacitance appears from this term.

 \section{Characterization of motile element}  
In the following, experimentally observable quantities are derived from the model so that values for the cellular parameters can be determined from experimental data.

\subsection{Load-free displacement}
Length changes of OHCs have been quantified by changing the membrane potential gradually or stepwise while measuring  cell length without load. That corresponds to describing $P_\infty$ as a function of $V$  under the condition of $K=0$. Thus, Eq.\ \ref{eq:distr} is accompanied by $\Delta G=q(V-V_{1/2})$ instead of Eq.\ \ref{eq:energy}. This equation has the same form as the one that has been used for fit experimental data~\cite{a1987}.

\subsection{Isometric force generation}
Isometric force generation per voltage changes can be obtained by evaluating $dF/dV$ from the equation $x=F/k+anP_\infty$ with Eq.\ \ref{eq:distr} and $\Delta G=[q(V-V_{1/2})-aF]$ for a given displacement $x$. This leads to,
\begin{equation} \label{eq:isoforce}
\frac{dF}{dV}=\frac{\gamma aqnk}{1+\gamma a^2nk}
\end{equation}
with $\gamma=\beta P_\infty(1-P_\infty)$. The dependence on length displacement $x$ enters through the value of $P_\infty$. The maximum value is $\beta aqnk/(4+\beta a^2nk)$ at $P_\infty=1/2$.

\subsection{Axial stiffness}
The effective compliance of the cell can be determined by $dx/dF$ for a given voltage $V$. If we introduce the effective stiffness $\tilde{k}$, this leads to,
\begin{equation} \label{eq:k}
\frac{1}{\tilde{k}}=\gamma a^2n+\frac 1 k
\end{equation}
Thus, the minimal value $\tilde{k}_\mathrm{min}$ of the effective stiffness  is $4k/(\beta a^2 nk+4)$ at $P_\infty=1/2$.

 \subsection{Nonlinear capacitance}\label{sec:cap}

Another characteristic quantity that described the motile element is a contribution of the motile element to the membrane capacitance, which is often referred to as nonlinear capacitance. Let us consider small periodic changes with amplitude $v$ in the membrane potential on top of a constant value $\bar{V}$,
\begin{align} \label{eq:V-period}
V(t) =\overline{V}+v \exp[i\omega t].
\end{align}
Then the response can be described by,
\begin{align}\label{eq:Pinf-period}
P_\infty(t) =&\overline{P}_\infty+p_\infty \exp[i\omega t], \\ \label{eq:P-period}
P(t) =&\overline{P}+p \exp[i\omega t],
\end{align}
where the variables expressed in lower case letters are small and those marked with bars on top are time-independent. 
Under time-independent condition, $\bar{P}=\bar{P_\infty}$ and $\bar{P}$ is expressed by Eq.\ \ref{eq:distr} with $\Delta G$, in which $P$ is replaced by $\bar{P}$.

If the amplitude $v$ is small, we can ignore second-order terms, Eqs.\ \ref{eq:distr} and  \ref{eq:motion} respectively lead to,
\begin{align} \label{eq:oscil-distr}
p_\infty=&-\beta\bar{P}(1-\bar{P})(qv+a^2n\tilde{K}p)\\ \label{eq:oscil-motion}
(-\omega^2m+i\omega\eta)p=&(k+K)(p_\infty-p). 
\end{align}
These equations lead to,
\begin{equation} \label{eq:p}
p=\frac{-\gamma q}{1+\gamma a^2 n \tilde{K}-(\omega/\omega_{r})^2+i\omega/\omega_{\eta}}\cdot v,
\end{equation}
with constants $\omega_{\eta}$, $\omega_{r}$, and $\gamma$, which respectively characterizes viscoelasticity, resonance, and the operating point: 
\begin{align} \nonumber
\omega_{\eta}&=(k+K)/\eta, \\ \nonumber
\omega_{r}^2&=(k+K)/m, \\ \nonumber
\gamma&=\beta\bar{P}(1-\bar{P}).
\end{align}

The capacitive current due to voltage changes is $i \omega q p \exp[i\omega t]$, which should be also expressed as $i\omega C_{nl}v\exp[i\omega t]$ using a component $C_{nl}$ of the membrane capacitance. Hence,
$C_{nl}=(qn /v) \Re[p]$ because conformational change $p$ of each motile element carries charge $p$ and the cell has $n$ such elements. The membrane capacitance $C_m$ is the sum of  $C_{nl}$ and the structural capacitance $C_0$, due primarily to lipid bilayer of the plasma membrane.  

In the following, the contribution to the capacitance is examined. The order is from the most restricted case, where this quantity is better studied, to more general cases.

\subsubsection{Mass-free and drag-free condition ($m=0$,  $\eta=0$)}
Let us start with the simplest case, where $m\rightarrow 0$ and $\eta\rightarrow 0$.  That leads to the total membrane capacitance $C_m$, expressed by
\begin{equation}
C_m=C_0+\frac{\gamma nq^2}{1+\gamma a^2n\tilde{K}},
\label{eq:capK}
\end{equation}
 
Letting $\tilde{K}\rightarrow 0$ or $a \rightarrow 0$, we recover the familiar expression for the membrane capacitance of outer hair cells $C_m=C_0+\gamma nq^2$ under load-free condition \cite{s1991,i2010}. In addition, this expression shows that nonlinear capacitance decreases with increasing elastic load.  That is consistent with intuition that constraints on the cell length reduce nonlinear capacitance. In an extreme limit, in which a rigid load does not allow transitions of the motor elements, nonlinear component diminishes. This expectation is consistent with greatly diminished nonlinear capacitance observed in rounded OHCs with constrained membrane area \cite{ai1999}.

\subsubsection{Mass-free condition ($m=0$) }
Now let us proceed to a more general case, in which the viscous term does not disappear. 
The total membrane capacitance $C_m$ is expressed,
\begin{equation}\label{eq:Cm@m=0}
C_m=C_0+\frac{\gamma nq^2(1+\gamma a^2n\tilde{K})}{(1+\gamma a^2n\tilde{K})^2+(\omega/\omega_{\eta})^2}.
\end{equation}
As expected, in the limit of low frequency, this expression turns into Eq.\ \ref{eq:capK}.  In the limit of $a \rightarrow 0$, it leads to the expression of the frequency dependence of nonlinear capacitance that was previously derived based on an assumption that transition rates between the states are intrinsic \cite{i1997}. The present interpretation supports the interpretation that the experimentally observed frequency roll-off \cite{ga1997,Dong2000} of the membrane capacitance is indeed the result of viscoelastic relaxation \cite{Dong2000,i2010}. It should be noted that the imaginary part of $p$ contributes to the conductance. However, it diminishes in both low frequency limit and high frequency limit.

\subsubsection{Mechanical resonance}
A similar evaluation of nonlinear capacitance can be performed for a system with non-zero mass, which has mechanical resonance. The quantity $p$ is expressed by,
\begin{equation}
C_m=C_0+\frac{\gamma n q^2[1+\gamma a^2 n \tilde{K}-(\omega/\omega_r)^2]}{[1+\gamma a^2 n \tilde{K}-(\omega/\omega_{r})^2]^2+(\omega/\omega_{\eta})^2},
\label{eq:Cm-mres}
\end{equation}

Under the condition $\omega_r\gg \omega$, Eq.\ \ref{eq:Cm@m=0} can be obtained. It should be noted that nonlinear capacitance disappears and $C_m=C_0$ at $\omega/\omega_r=1+\gamma a^2n\tilde{K}$. In addition, the capacitance    becomes quite singular near this resonance frequency as the characteristic frequency  $\omega_{\eta}$ of viscoelasticity exceeds the resonance frequency $\omega_r$.   A reduction in the capacitance $C_m$ near resonance leads to an increase in the receptor potential because it reduces the attenuation due to the resistance-capacitance (RC) circuit in the cell. That is analogous to piezoelectric resonance.

\begin{figure}[h]
\begin{center}
 
\includegraphics[width=0.5\textwidth]{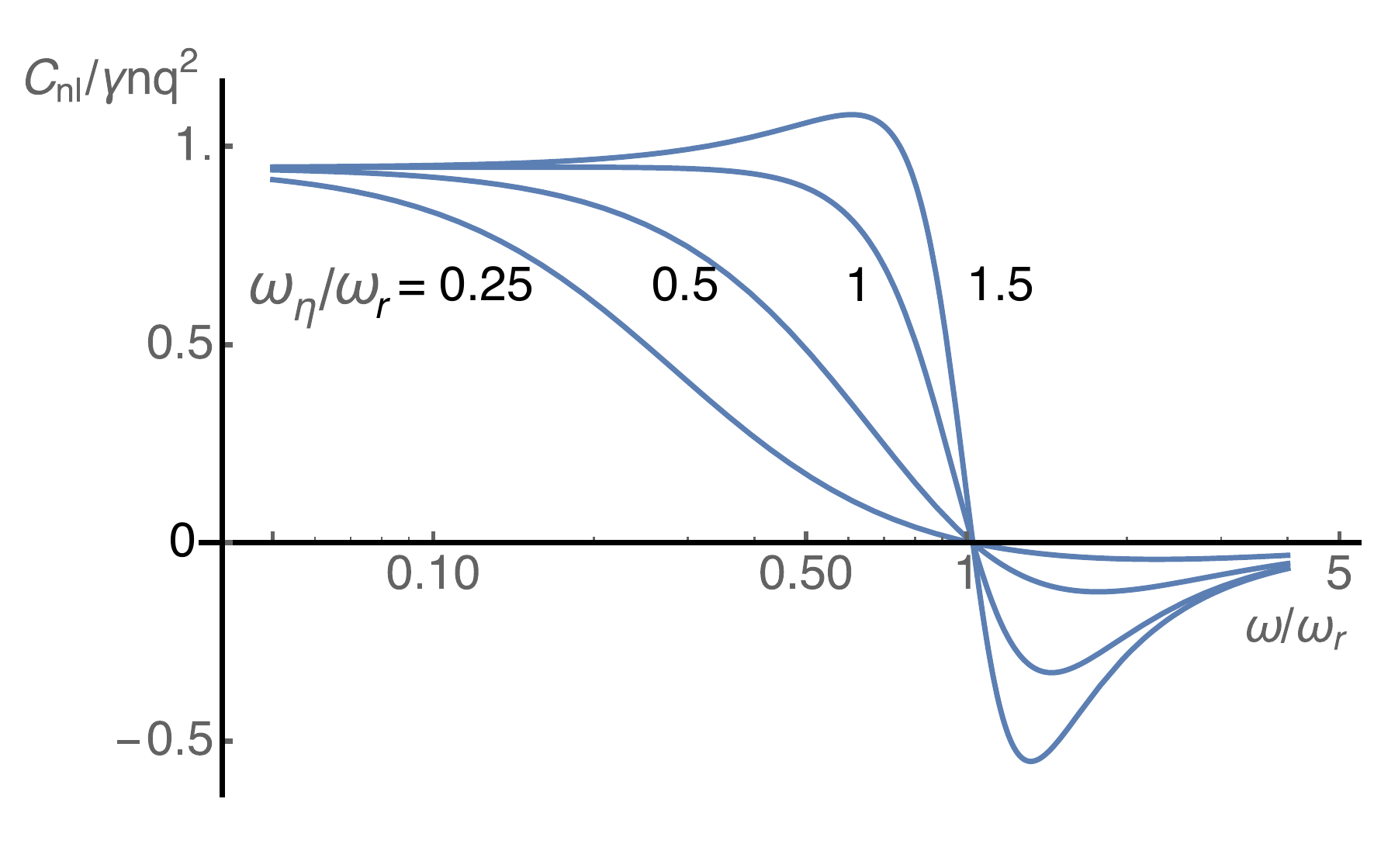}
\caption{\small{Contribution $C_{nl}$ to the membrane capacitance in the presence of mechanical resonance. The capacitance normalized to its maximum value $\gamma nq^2$ is plotted against the ratio of the frequency $\omega$ to the resonance frequency $\omega_r$. The plots correspond respectively to $\omega_\eta/\omega_r= 0.25, 0.5, 1, \mathrm{and}\; 1.5$ by changing either the drag coefficient $\eta$ or the mass $m$, while keeping the elastic load $K$ constant. The value for $\gamma nq^2\tilde{K}$ is assumed to be 0.05.}}
\label{fig:cnl}
 
\end{center}
\end{figure}

\section{Parameter values}\label{sec:param} 

\begin{table}[h] 
\begin{center}
\begin{tabular}{c|r|r|l}
\hline\hline
parameter & experimental & used (unit) & remarks \\
\hline
$e_{ec}$ & $\sim 90$ & 90 (mV)& $^a$\\
$e_{K}$ & $-90$ & $-90$ (mV)& $^a$\\
$C_0$ & $ 12$  & $12$ (pF)&$^b$\\
$R_m$ & $ 10$  & $10$ (M$\Omega$)&$^c$\\
$R_a$ & $ \sim40$  & $42$ (M$\Omega$)&$^c$\\
\hline\hline
$q$ & $\sim 0.8 $  &  $0.8 $ ($e$)&$^d$\\
$n$ & $\sim 15 \times 10^6$ & $15 \times 10^6$& $^d$\\
$an$ & $1.2\pm0.1$ & 0.96 ($\mu$m)& $^e$\\
$a$ & &$6.4 \times 10^{-5}$ (nm)&  \\
$k$ & & $0.016$ (N/m)& \\
\hline
& & predicted &\\
\hline
$dF/dV$ & $0.1\pm0.04$  & 0.11 (nN/V) &$^f$\\
$\tilde{k}$ & $0.017\pm0.005$ & $0.015 $ (N/m)&$^f$\\
\hline
\end{tabular}
\end{center}
\caption{A set of parameter values for outer hair cells with $12$ pF linear capacitance $C_0$. These values corresponds to OHCs from the region with best frequency of 4 kHz for rats and gerbils \cite{Johnson2011}.   The quantities in the upper rows are electric properties and lower rows characterize the motile element.  
The values for resistances are from rats and values of motility-related parameters are from guinea pigs.
$^a$ The sum of the endocochlear potential ($e_{ec} \sim$90 mV) \cite{salt1989} and the reversal potential ($e_K$) of the basolateral membrane ($\sim -90$ mV). 
$^b$ \cite{skkkt1998}.
$^c$ \cite{Johnson2011}.
$^d$ The unit mobile charge $q$ (in the electronic charge $e$), and its number $n$ have been determined by nonlinear component of the membrane capacitance.
$^e$ The amplitude of load free displacement $an$ is between 4 and 5\% of the total length.
$^f$ Both force generation and the elastic modulus do not show length dependence \cite{ia1997}. The stiffness value corresponds the elastic modulus of 0.51 $\mu$N per unit strain for a 30 $\mu$m-long cell.}
\label{table:paramvals}
\end{table}

  Now we determine the parameters using experimental data. Those quantities   important for determining the parameters are: the amplitude $an$ of the load-free mechanical displacement, which is between 4 and 5 \% of the cell length, the steepness of load-free mechanical displacement and that of nonlinear capacitance, both of which are characterized by $q$, and the number $n$ of the motile elements in the cell, which is obtained by dividing the total charge movement $Q$ divided by $q$. In addition, experimental values are available for the maximal value of nonlinear capacitance $\beta n q^2/4$, the axial stiffness of the cell $\tilde{k}$, and isometric force production $dF/dV$. The parameter values are listed in Table \ref{table:paramvals}.

For numerical examination, we assume that the cell length of the OHC is 30 $\mu$m and that the motor is 40 \% in the elongated state in the natural length and the receptor potential is generated by 10\% change in the resting hair bundle resistance $R_a$, which corresponds to the condition where 30\% of the mechanotransducer channels are open.  

Here we now examine the consistency with an assumption, which we made for deriving the equation of motion (Eq.\ \ref{eq:motion}). Recall that the minimal value of the stiffness of the motile element is $4/(\beta a^2n)$ at $P_\infty= 1/2$ (Eq.\ \ref{eq:k}). For the present set of the parameter values, the value for this stiffness is $0.27$ N/m, larger than $0.016$ N/m for the intrinsic stiffness $k$. Thus this set of parameter values is consistent with the assumption made earlier. 

The consistency of the one-dimensional model is tested by comparing its predicted values and experimental values for force generation $dF/dV$ and the stiffness $\tilde{k}$. The predicted values are within experimental errors (Table \ref{table:paramvals}).

 \section{Response to small periodic stimulation}
Here we assume small periodic changes with an angular frequency $\omega\;(=2\pi f)$ from a resting resistance $\bar{R}_a$ of the hair bundle resistance, 
\begin{equation*}
\label{eq:ra} R_a(t)=\bar{R}_{a}+r \exp[i\omega t],
\end{equation*}
leads to the receptor potential $V(t)=\overline V+v\exp[i\omega t]$. Eq.\ \ref{eq:potential} leads to
\begin{eqnarray}
\label{eq:oscil-voltage}
-\frac{e_{ec}-\bar{V}}{\bar{R_a}}\frac{r}{\bar{R_a}}=\left(\frac 1 {\bar{R}_a}+\frac 1 {R_m}\right)v+i\omega(C_0v-nq\cdot p),
\end{eqnarray}
with $\bar{V}=(e_{ec}R_m+e_K\bar{R}_a)/(R_m+\bar{R}_a)$. The amplitude $v$ of the potential leads to motile responses as described by $p$ and $p_\infty$ in Eqs.\ \ref{eq:Pinf-period} -- \ref{eq:oscil-motion}.

The receptor potential with amplitude $v$ elicited by small periodic changes in the hair bundle resistance induces changes in the amplitude $p$ of changes in the fraction of the extended state of the motile element, resulting in displacements and force generation of the cell. Because of mechanoelectric coupling, the load on the motile element reciprocally affects $p$, which, in turn, attenuates the receptor potential $v$.  

\section{Examination of energy output}
From here on,   energy balance is examined for small periodic stimulation at the hair bundle  in a mass-free system ($m=0$) using the parameter values in Table \ref{table:paramvals},   which corresponds to a best frequency of 4 kHz. 

Non-zero mass introduces an additional time constant, requiring examining a larger number of cases. In addition, the mass may depends on the frequency because it may include fluid mass \cite{fw1988}, and hair cell mass may not constitute a major part of the mass \emph{in vivo}.

\subsection{Dependence on elastic load}\label{sec:elastic_load} 

  Before evaluating energy, it is useful to examine the membrane capacitance, which is the attenuating factor of the receptor potential.   
An external elastic load $K$ affects the membrane capacitance through $ \gamma a^2 n \tilde{K}$ (See Eqs.\ \ref{eq:capK} and \ref{eq:Cm@m=0}). Since $\tilde{K}(=kK/(k+K))\rightarrow k$  for $K\rightarrow \infty$, the maximum value of this factor is $(1/4) \beta a^2 nk$ at $\bar{P}=1/2$. For our set of parameter values, the maximum value of this factor is 0.11, indicating that this factor is rather small even though it is not negligible at low frequencies ($\omega\ll\omega_{\eta}$). For $\omega\gg\omega_{\eta}$, nonlinear capacitance diminishes and so does the effect of an elastic load on the membrane capacitance. 

In the range where $\omega$ is comparable to the viscoelastic characteristic frequency $\omega_{\eta}$, however, the effect of the elastic load $K$ appears mainly through the characteristic viscoelastic frequency $\omega_{\eta}(=\tilde{K}/\eta)$. For a given frequency $\omega$, an increase in the elastic load $K$ increases $\omega_{\eta}$ through $\tilde{K}$ and thus increases the capacitance $C_m$.

\subsubsection{Energy output}
In the following, energy output from OHC is examined assuming that the hair bundle is stimulated at a level at which hair bundle conductance undergoes changes with an amplitude 10\% of the resting value, where the mechano-channels are assumed to be 30\% open. Thus, the change is 3\% of its maximal value. Parameter values are given in Table \ref{table:paramvals}. This amplitude allows linearized approximation, Eqs. \ref{eq:oscil-distr}, \ref{eq:oscil-motion}, and \ref{eq:oscil-voltage}. In this regime, energy output increases with the second power of the receptor potential.

The system has two characteristic frequencies,   the frequency $\omega_\eta$ of viscoelastic roll-off and the roll-off frequency $\omega_{RC}(\approx 1/R_mC_m)$ of the RC circuit.   Here the effect of the elastic load $K$ is examined.  The stiffness $k$ of the cell has been determined by experiments (Table \ref{table:paramvals}). Two characteristic frequencies of the system leads to two cases, which are examined here: $\omega_\eta \gg \omega_{RC}$ and $\omega_\eta \sim \omega_{RC}$. It should be noticed, while $\omega_{RC}$ is independent of $K$, $\omega_\eta$ goes up as $K$ increases. The  energy $E_e$ conveyed to the external elastic load $K$ per half cycle can be obtained by evaluating $(1/2) K |x|^2=(1/2) K \cdot |(\tilde{K}/K) na p|^2$ and the work $E_d$  against viscous drag per half cycle is $(1/2)\eta\omega|x|^2=(1/2)\eta\omega\cdot |(\tilde{K}/K) na p|^2$.

\subsubsection{small viscous drag ($\omega\ll\omega_{\eta}$)}\label{subsubsection:small_eta}
\begin{figure}[h!] 
\begin{center} 
\begin{minipage}[b]{6cm}\textbf{A}  \end{minipage}  \begin{minipage}[b]{6cm} \textbf{B}  \end{minipage}\\
\includegraphics[width=6cm]{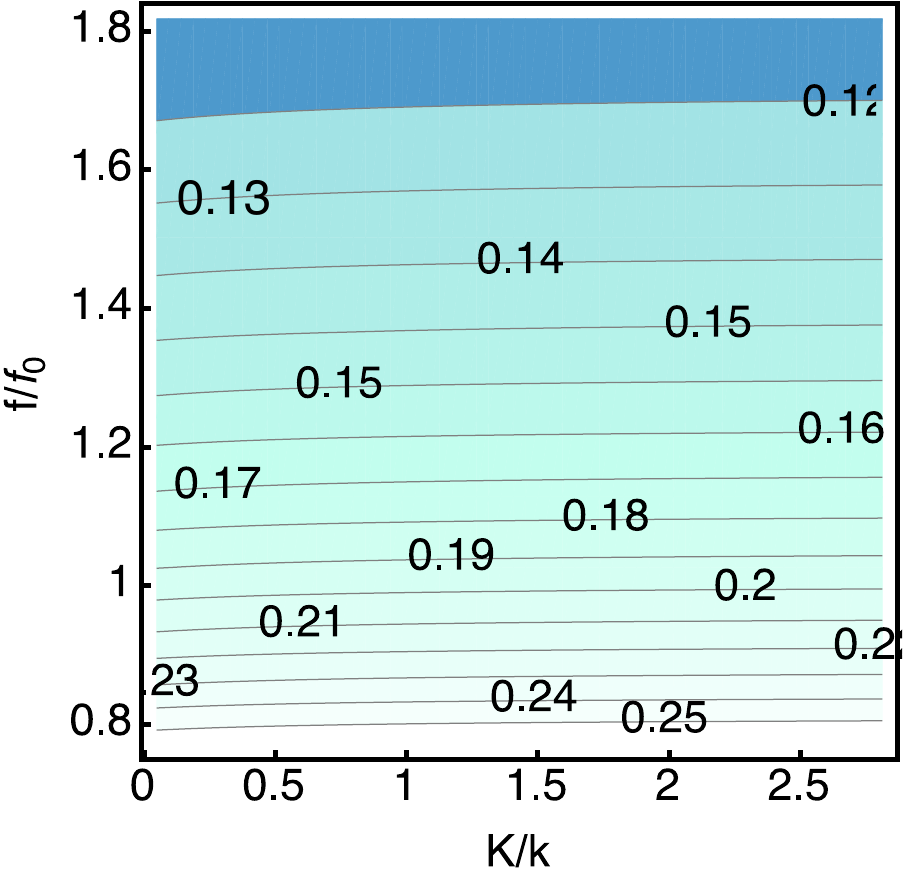} \ \ \includegraphics[width=6cm]{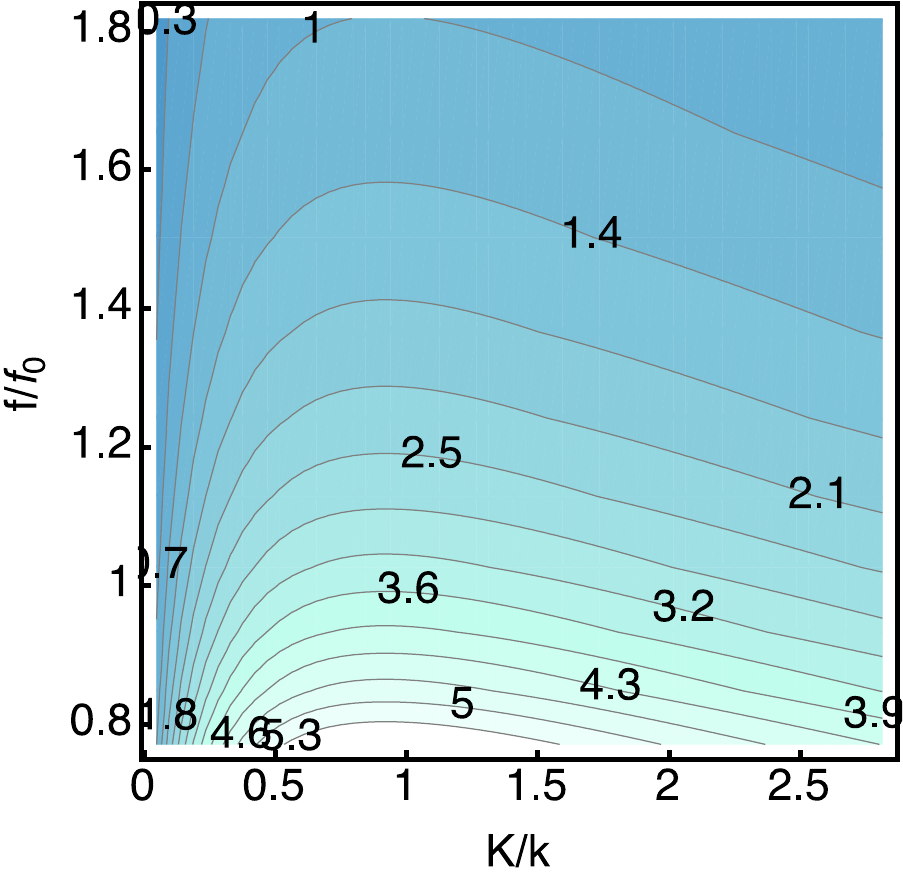} 
\caption{\small{Small viscous drag ($\omega_\eta \gg \omega_{RC}$). The dependences on the elastic load (the horizontal axis) and the frequency (the vertical axis) are shown as color coded contour plots. Elastic load is described by the ratio $K/k$, where $k$ is the intrinsic stiffness of the cell. Frequency is represented by the ratio $f/f_0$, with $f_0=$4 kHz, which corresponds to the parameter values of the cell given in Table \ref{table:paramvals}.
\textbf{A}: The amplitude of the receptor potential.  The values (in mV) are given in the plot.
\textbf{B}: The work against elastic load during a half cycle. The values  (in zJ =$10^{-21}$}J) are given in the plot. $\eta=10^{-10}$ kg/s.}
\label{fig:small_eta}
\end{center} 
\end{figure}

Here examine a case in which the drag coefficient $\eta$ is extremely small and viscous loss is negligible. Let $\eta=10^{-10}$ kg/s. This value satisfies the condition $\omega_\eta \gg \omega_{RC}$ except for diminishing $K$. This condition is satisfied even with increasing external elastic load $K$ to $\sim10k$. At a given frequency, an increase in the amplitude of the receptor potential with increasing external elastic load $K$ is too small to notice in the plot (Fig.\ \ref{fig:small_eta}A). With a given elastic load, the amplitude of the receptor potential monotonicaly decreases with increasing frequency (Fig.\ \ref{fig:small_eta}A).

The work against the elastic load is evaluated for a half cycle. For a given frequency, elastic energy output per half cycle has a maximum with respect to the external elastic load $K$ at the load ratio $K/k\approx1$ (Fig.\ \ref{fig:small_eta}B). The optimal ratio is not significantly affected by the frequency. Work output monotonically decreases with the frequency (Fig.\ \ref{fig:small_eta}B). The plot also shows that decrease of energy with frequency is less steep at larger load.
 Asymmetry increases with increasing frequency, while the peak ratio $K/k$ remains virtually unchanged.  

\subsubsection{larger viscous drag ($\omega\sim\omega_{\eta}$)}
\begin{figure}[h!] 
\begin{center} 
\begin{minipage}[b]{6cm}\textbf{A}  \end{minipage}  \begin{minipage}[b]{6cm} \textbf{B}  \end{minipage}\\
\includegraphics[width=6cm]{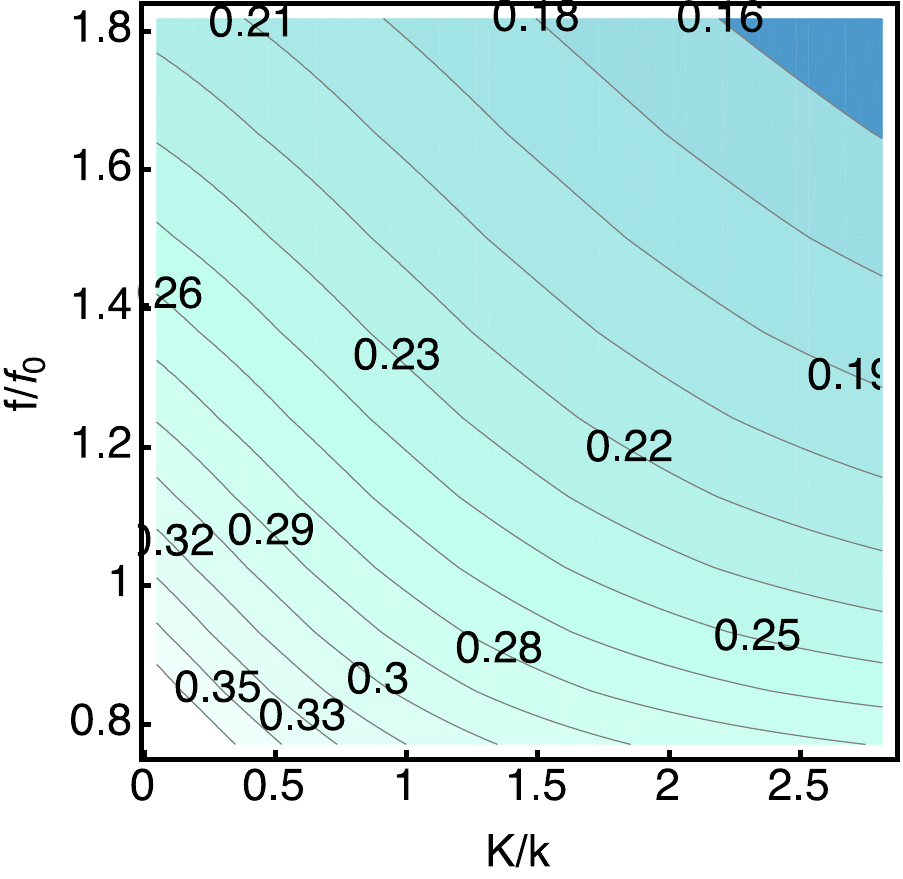}\ \ \includegraphics[width=6cm]{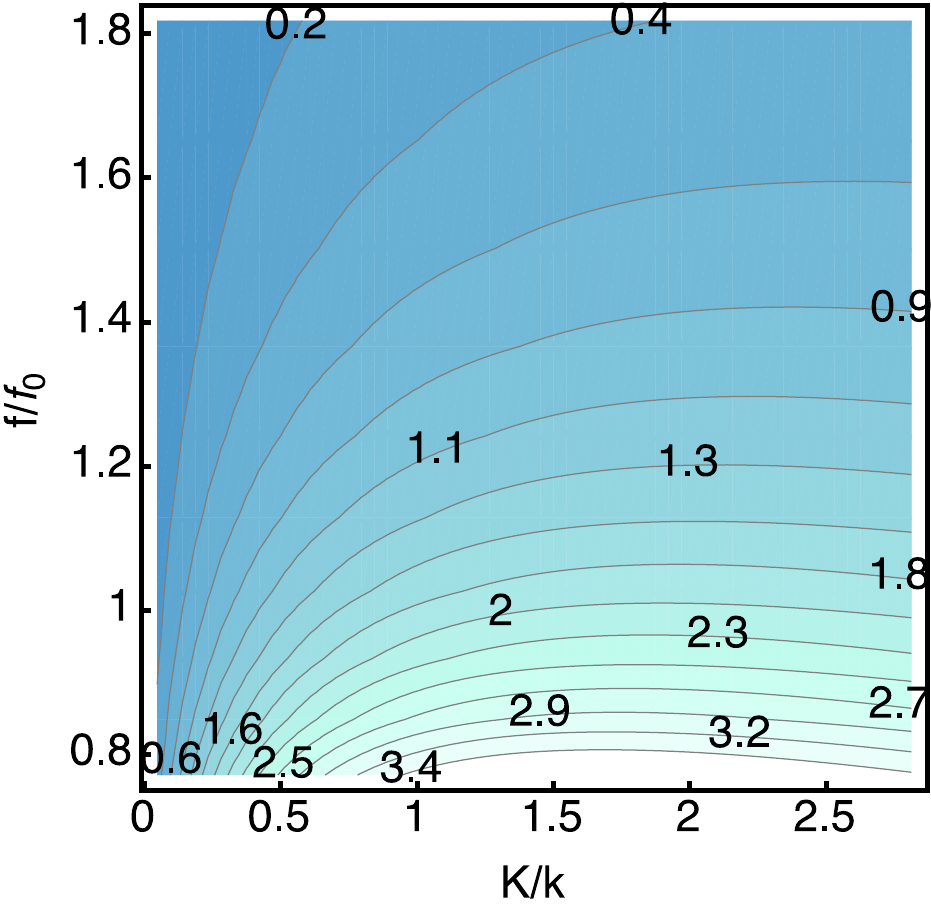}\\
\begin{minipage}[b]{6cm}\textbf{C}  \end{minipage}  \begin{minipage}[b]{6cm} \textbf{D}  \end{minipage}\\
\includegraphics[width=6cm]{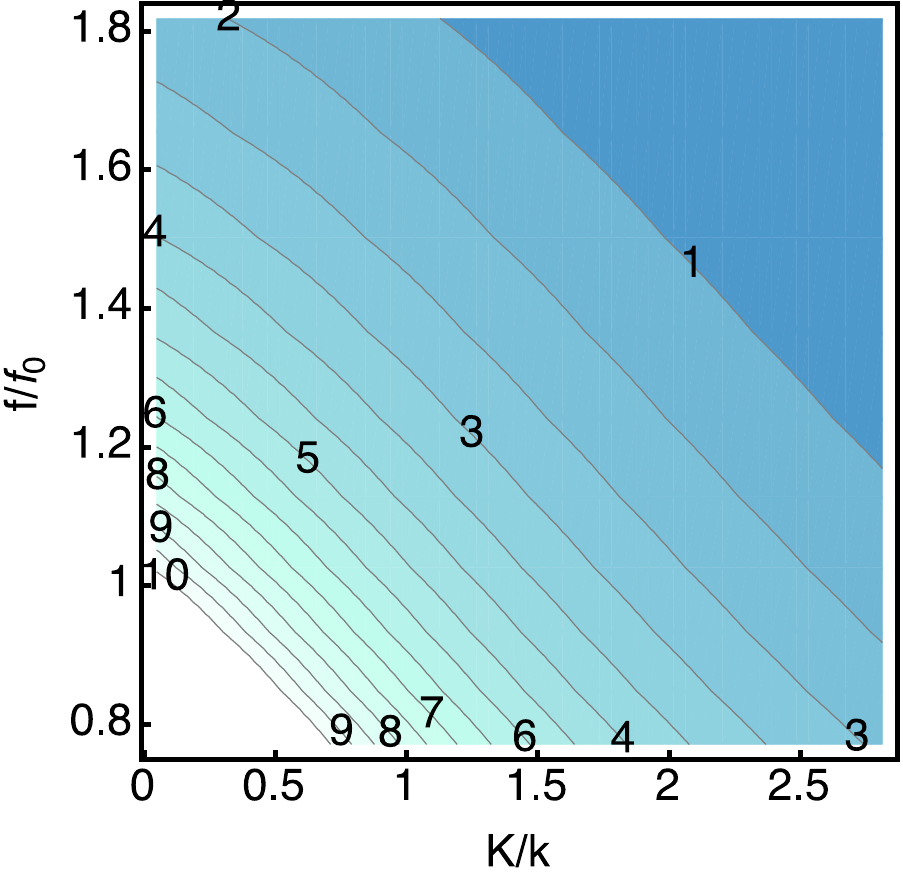}\ \ \includegraphics[width=6cm]{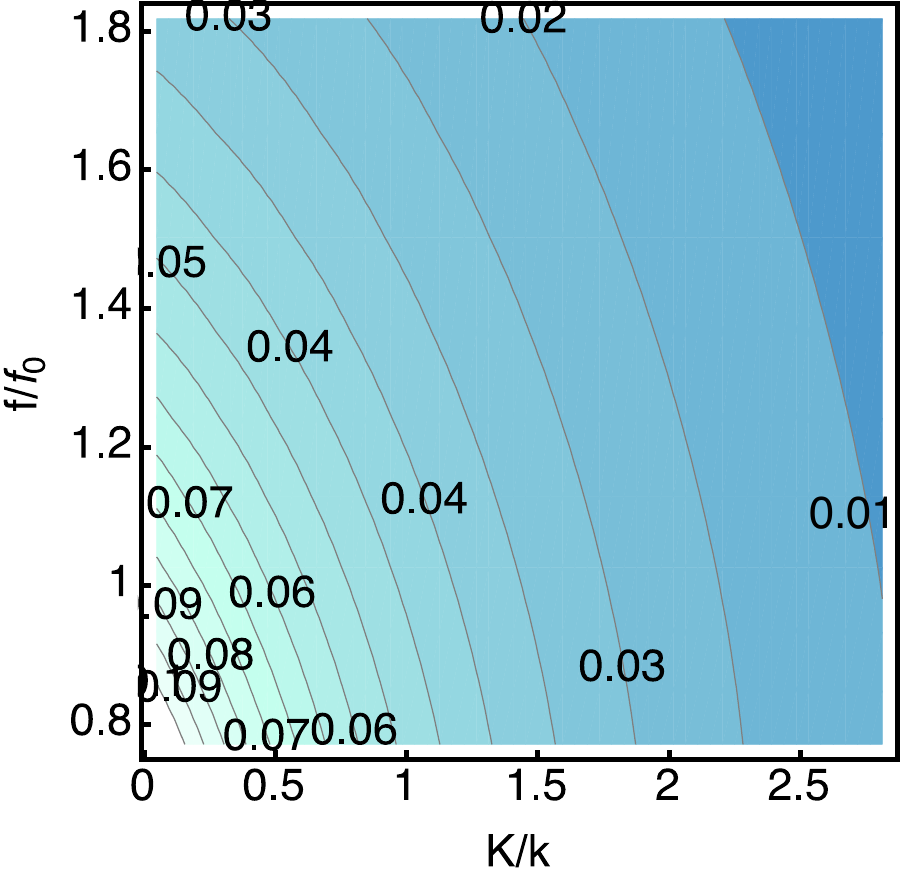}\\
\caption{\small{ The effect of elastic load for a larger drag ($\omega_\eta \sim \omega_{RC}$). The dependences on the elastic load (the horizontal axis) and the frequency (the vertical axis) are shown as color coded contour plots. Elastic load is described by the ratio $K/k$, where $k$ is the intrinsic stiffness of the cell. Frequency is represented by the ratio $f/f_0$, with $f_0=$4 kHz, which corresponds to the parameter values of the cell given in Table \ref{table:paramvals}.
\textbf{A}: Amplitude (in mV) of the receptor potential.
\textbf{B}: The work against the elastic load during a half cycle. The values are given in zJ (=$10^{-21}$J). 
\textbf{C}: The work (in zJ) against the viscous load during a half cycle. 
\textbf{D}: Power output in fW (=$10^{-15}$W) working against the viscous load.
$\eta=2\times 10^{-6}$ kg/s.}}
\label{fig:large_eta}
\end{center} 
\end{figure}

 If the viscous drag is larger, the receptor potential decreases with increasing the stiffness ratio $K/k$ for a given frequency (Fig.\ \ref{fig:large_eta}A). This result may appear counterintuitive because it is the opposite an increase, be it rather small, under the condition of small drag. This reversal is due to a change in the time constant. The membrane capacitance increases with increasing elastic load, owing to decreasing $\omega/\omega_{\eta}$ in Eq.\ \ref{eq:Cm@m=0}. An increase in the membrane capacitance increases RC attenuation, resulting in a reduction of the receptor potential. The frequency dependence of the receptor potential is monotonic. However, smallest elastic load makes the slope steeper at higher frequencies (Fig.\ \ref{fig:large_eta}A).

The value of $2\times 10^{-6}$kg/s for the drag coefficient used in the plots is chosen to examine a condition, which nearly maximizes the work against viscous drag, as we will see later.
For each given frequency, the work against the elastic load has a maximum with respect to the stiffness ratio (Fig.\ \ref{fig:large_eta}B). The load ratio $K/k$ that maximizes the work is close to two for low frequencies, unlike with lower viscous drag. In addition, for higher frequencies the ratio that maximizes energy output increases significantly (Fig.\ \ref{fig:large_eta}B). The maximal value at the frequency $f_0$(=4kHz) is about 2 zJ (=$2\times 10^{-21}$ J) at $K/k=1.5$.

The work against viscous drag per half cycle decreases with increasing frequency (Fig.\ \ref{fig:large_eta}C). It also decreases monotonically with increasing stiffness ratio $K/k$ unlike the work against the elastic load. For this reason, the comparison of magnitude with elastic work heavily depends on the ratio $K/k$. The work against viscous drag is much larger than the work against elastic load if the elastic load is small. At $K/k=1.5$, the value at the frequency $f_0$ is 6 zJ, still larger than than the value 2 zJ for the elastic work.

Power output can be obtained by multiplying the work against drag per half cycle by twice the frequency $2f$. It is less frequency dependent, even though it still decreases with increasing frequency (Fig.\ \ref{fig:large_eta}D). 

\subsection{Dependence on viscous load}
Here we examine the effect of the viscous load for fixed values of the elastic load.  Two values of the external load would be of interest: One of them is the case in which the stiffness of the external elastic load is similar to the internal stiffness of the cell. The other is the case without an external elastic load. The former condition is presumably close to the physiological condition and the output in the form of elastic energy would be appreciable. The latter case is also of interest because it provides the maximal work against the viscous drag. Here we use $\eta_0=10^{-6}$ kg/s as the unit of drag coefficient.

\subsubsection{With elastic load ($K=k$)}
\begin{figure}[h!]
\begin{center}
\begin{minipage}[b]{6cm}\textbf{A}  \end{minipage}  \begin{minipage}[b]{5cm} \textbf{B}  \end{minipage}\\
\includegraphics[width=6cm]{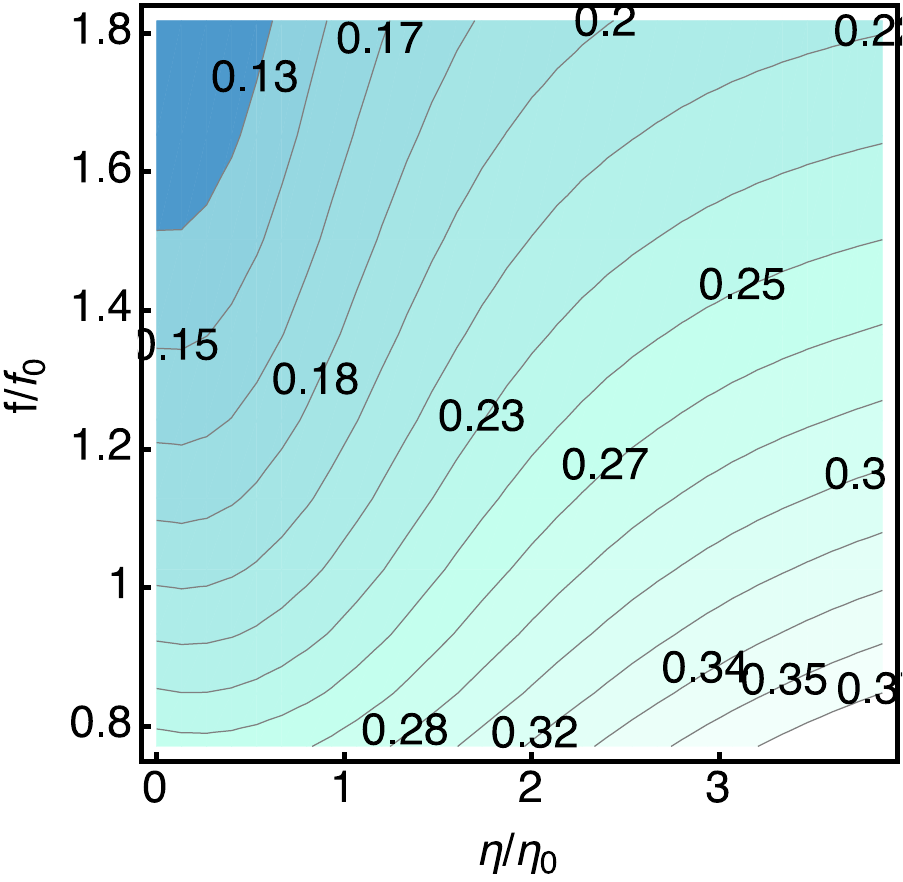} \ \ \includegraphics[width=6cm]{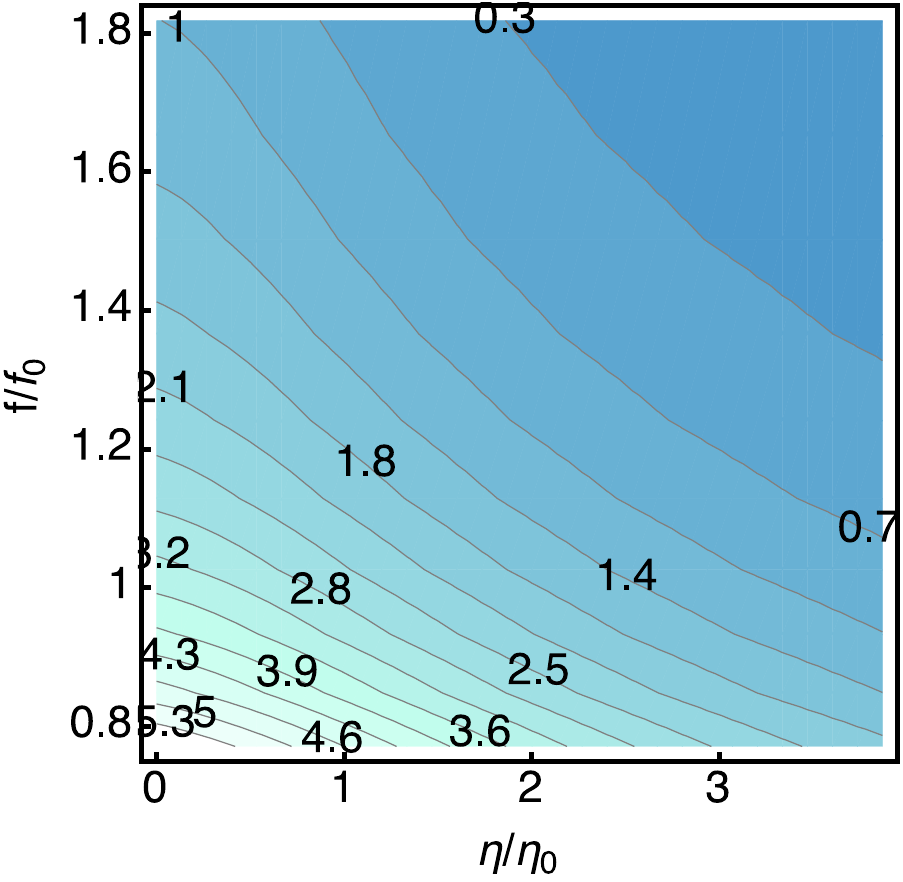}
\begin{minipage}[b]{6cm}\textbf{C}  \end{minipage}  \begin{minipage}[b]{5cm} \textbf{D}  \end{minipage}\\
\includegraphics[width=6cm]{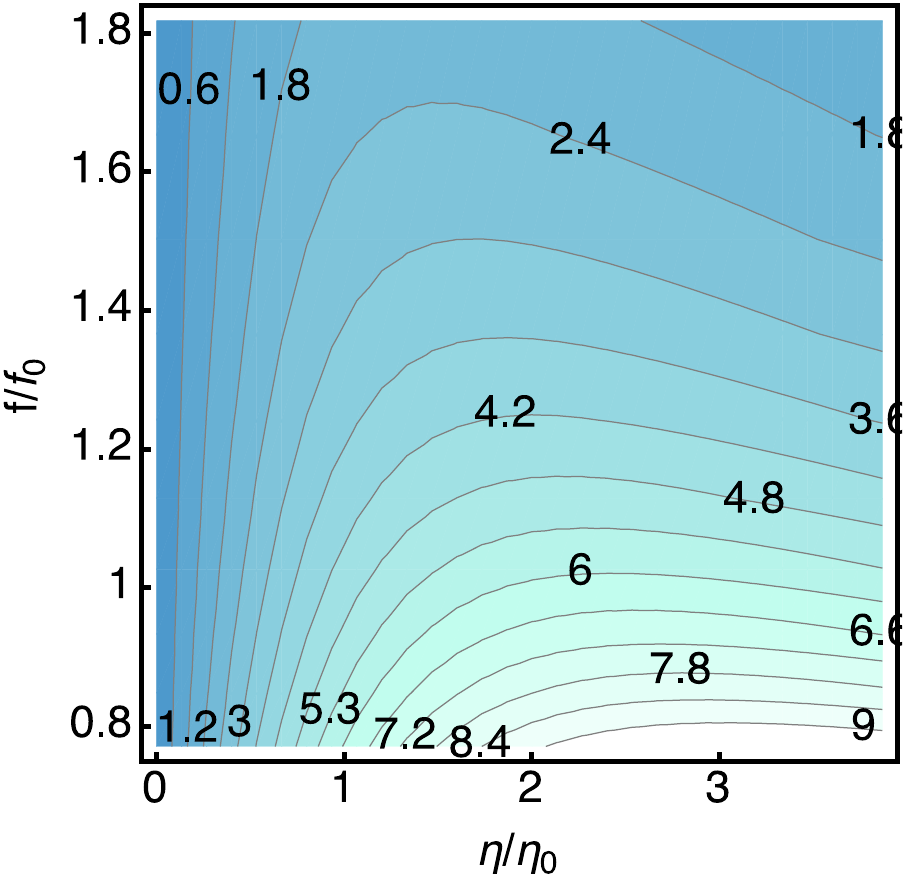} \ \ \includegraphics[width=6cm]{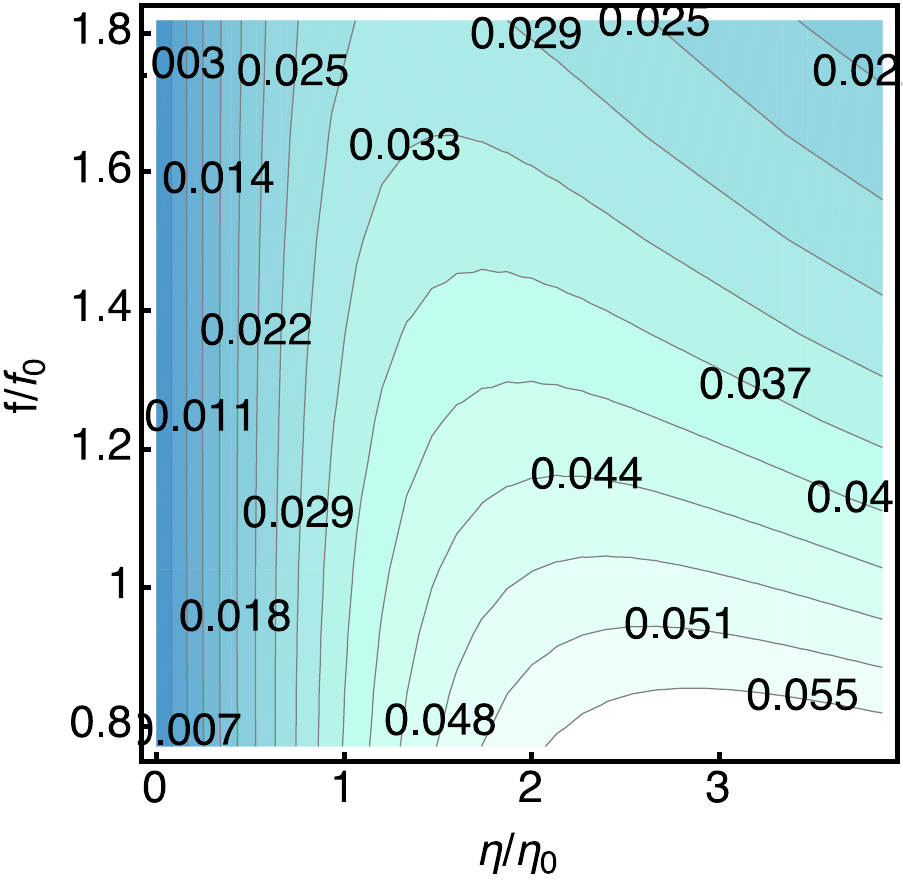}
\end{center}
\caption{\small{The effect of drag in the presence of elastic load $K(=k)$. The dependences on drag (the horizontal axis) and the frequency (the vertical axis) are shown as color coded contour plots. Drag is described by the ratio $\eta/\eta_0$, where $\eta_0=1.0\times 10^{-6}$ kg/s. Frequency is represented by the ratio $f/f_0$, with $f_0=$4 kHz, which corresponds to the parameter values of the cell given in Table \ref{table:paramvals}.
\textbf{A:} The amplitude (in mV) of the receptor potential.
\textbf{B:} The work (in zJ=$10^{-21}$J) against the elastic load during a half cycle.
\textbf{C:} The work (in zJ) against the viscous drag during a half cycle.
\textbf{D:} Power output (in fW=$10^{-15}$W) working against the viscous drag.
}}
\label{fig:drag_effect}
\end{figure}

The receptor potential significantly increases with increasing viscous drag (Fig.\ \ref{fig:drag_effect}A). A higher drag coefficient leads to less steep decline of the membrane potential with increasing frequency in the middle range of the plot  (Fig.\ \ref{fig:drag_effect}A). 

The work against the elastic load per half cycle decreases with increasing drag as well as frequency  (Fig.\ \ref{fig:drag_effect}B), as intuitively expected. At frequency $f_0$, it is up to $\sim$3.4 zJ for low drag. At $\eta=2\eta_0$, the value is about 1.8 zJ.

The work against the viscous drag per half cycle monotonically decreases with the frequency. However, it has a maximum with respect to viscosity at a given frequency (Fig.\ \ref{fig:drag_effect}C). At $f_0$, the maximal value is  about 6 zJ at $\eta=2\eta_0$.

Power output due to the work against viscous drag (obtained by multiplying the dissipative energy output per half cycle by $2f$, twice the frequency) has a less steep frequency dependence (Fig.\ \ref{fig:drag_effect}D).  The maximal output at the frequency $f_0$ is 0.05 fW.

\subsubsection{Without elastic load}
At a given frequency, work against viscous drag is maximal at $K=0$, i.e. in the absence of an external elastic load (Fig.\ \ref{fig:large_eta}). For this reason it is interesting to examine the system without external elastic load to evaluate the limit even though such a condition may not be physiological.

\begin{figure}[h!]
\begin{center}
\begin{minipage}[b]{6cm}\textbf{A}  \end{minipage}\\  
\includegraphics[width=6cm]{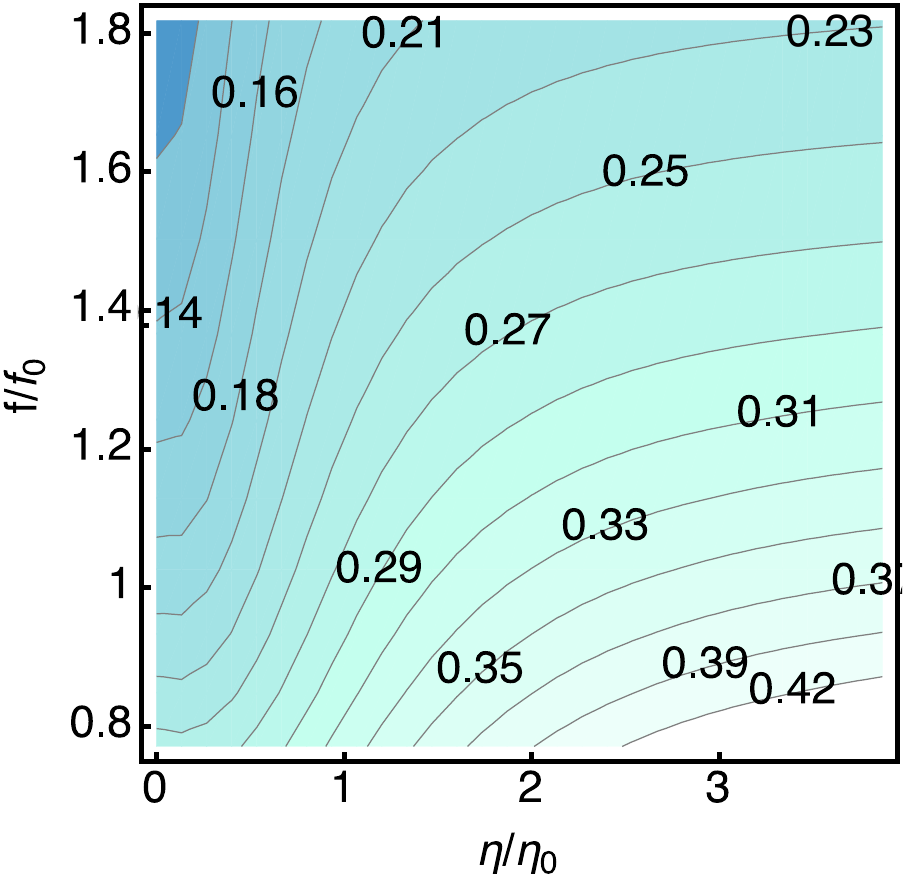}\\
\begin{minipage}[b]{5cm} \textbf{B}  \end{minipage}\ \ \begin{minipage}[b]{6cm}\textbf{C}  \end{minipage}  \\
\includegraphics[width=6cm]{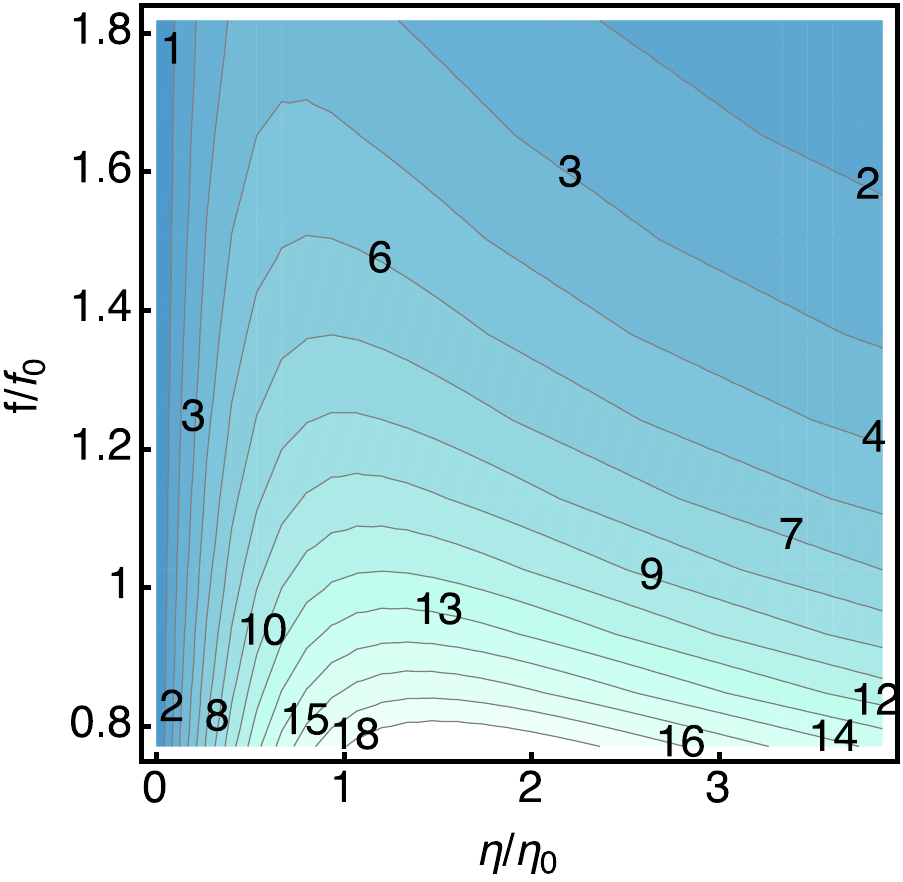}\ \ \includegraphics[width=6cm]{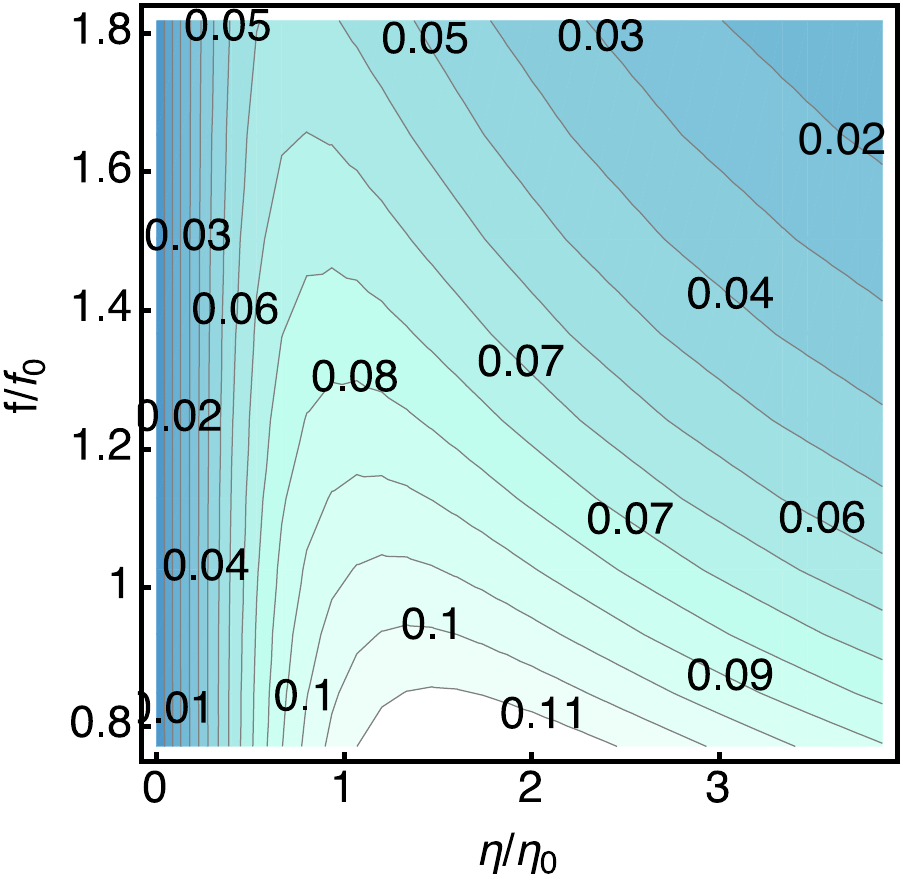} 
\end{center}
\caption{\small{The effect of drag in the absence of elastic load. The dependences on drag (the horizontal axis) and the frequency (the vertical axis) are shown as color coded contour plots. Drag is described by the ratio $\eta/\eta_0$, where $\eta_0=1.0\times 10^{-6}$ kg/s. Frequency is represented by the ratio $f/f_0$, with $f_0=$4 kHz, which corresponds to the parameter values of the cell given in Table \ref{table:paramvals}.
\textbf{A:} The amplitude (in mV) of the receptor potential.
\textbf{B:} The work (in zJ$=10^{-21}$J) against viscous drag during a half cycle.
\textbf{C:} Power output (in fW$=10^{-15}$W) working against the viscous drag.
}}
\label{fig:drag_only}
\end{figure}

The receptor potential increases with increasing viscous load and decreases with increasing frequency (Fig.\ \ref{fig:drag_only}A). At a given frequency, the work against the viscous drag per half cycle has a maximum (Fig.\ \ref{fig:drag_only}B). The peak position shifts to at smaller drag with increasing frequency. The maximal work per half cycle is about 12 zJ for $f_0$(=4kHz). The power output shows dependence on the drag and frequency similar to the work per half cycle does, even though the frequency dependence is less steep  (Fig.\ \ref{fig:drag_only}C). The maximal power output is about 0.1 fW.

\section{Discussion}\label{sec:disc}
  Here we discuss energy balance in an OHC that operates at 4 kHz in the absence of mechanical resonance by examining the elastic load and viscous drag, to which the cell is likely subjected. That is followed by discussion on cells that operates at higher frequencies. 

  \subsection{Internal drag}\label{subsection:loads}
Of the internal mechanical load, there is no need to discuss the stiffness of the OHC because it has been determined experimentally.  Some discussion would be needed regarding the internal drag of an OHC.  For that purpose, the cell body is approximated by a cylinder.    The velocity of the fluid is null in the middle of the cell and linearly increases towards the two ends, exactly the same as the plasma membrane.  For this reason, Poiseuille's law, for example, does not apply and viscous drag must be very small. 

The magnitude of the drag on the outer surface could be estimated by Stokes' law for a sphere, $6\pi\mu Rv$, where $R$ is the radius. The resulting drag coefficient is $6.6\times 10^{-8}$ kg/s, assuming $R=5\mu$m and  $0.7\times 10^{-3}$ kg/(m$\cdot$s) of water for the viscosity $\mu$. Since this value is smaller than the value $\sim10^{-6}$ that maximizes dissipative energy output, the assumed drag coefficient $\eta$ in Fig.\ \ref{fig:large_eta} and $\eta_0$ in Figs.\ \ref{fig:drag_effect} and \ref{fig:drag_only}, are dominated by the external viscous load and are not intrinsic to the cell.

\subsection{Receptor potential}
The numerical examination shows that the receptor potential is affected by an increase of elastic load and that of drag quite differently. While an increase in drag always increase the receptor potential (Fig.\ \ref{fig:drag_effect}A), an increase in the elastic load increases the receptor potential only slightly if drag is small (Fig.\ \ref{fig:small_eta}A) but it decreases the receptor potential if drag is larger (Fig.\ \ref{fig:large_eta}A). These observations suggest that drag is more effective than elastic load in enhancing the receptor potential that powers the cell's motility. 

\subsection{Amplifier gain}
The functional significance of OHCs as the cochlear amplifier depends on the balance between energy input and output. Energy output from the cell body has been evaluated in earlier sections. Energy input required for stimulating the hair bundle to generate the receptor potential consists of two components. One is elastic and the other is dissipative. The elastic component is recovered in the next cycle of stimulation during sustained oscillation. The work against viscous drag cannot be recovered. Therefore the balance of this energy has been considered critical for the sensitivity and sharp tuning of the ear \cite{g1948}. 

The present model allows an evaluation of OHC's output, which compensates for the dissipated energy for hair bundle stimulation. Examination of force or energy balance in the cochlea, however, cannot be made without a certain set of assumptions including the mode of motion in the cochlea \cite{odi2003a,Ramamoorthy2012a}. 

In the following, energy balance of a single OHC is examined. For this comparison, output values obtained under the condition $K\approx k$ and $\eta \approx 2 \eta_0$ are used so that the two kinds of outputs can be realized at the same time.

\subsubsection{Amplifier gain -- dissipative energy}
The hair bundle drag, which gives rise to the dissipative energy required for input has been evaluated for bull frog saccular hair bundles \cite{KozlovBaumgartRislerEtAl2011,Bormuth2014}.  It consists of external friction between the hair bundle and the bulk fluid and internal friction due to relative motion between the stereocilia. The former is dominant when a hair bundle is stimulated by force applied to the kinocilium: the drag coefficient of a whole hair bundle is about 1/5 of a single stereocilia \cite{KozlovBaumgartRislerEtAl2011}. If we consider a sphere with a diameter of 8.5 $\mu$m, which is the hair bundle height, Stokes' law gives the drag coefficient of 80 nNs/m, about the same as the drag coefficient calculated for the bundle in the frequency range higher than 2 kHz, where stereocilia moves in synchrony in response to force applied to the kinocilium \cite{KozlovBaumgartRislerEtAl2011,Bormuth2014}. This value can be even larger if the thickness of boundary layer is   added to the diameter.   Therefore, the internal drag must be smaller than the total drag at least by an order of magnitude. Here we set 1/10 of the total friction coefficient as an upper bound of the contribution of internal friction.

In the physiological conditions for OHCs, unlike a frog hair bundle stimulated by holding kinocilium, the hair bundle is in the subtectorial space between the tectorial membrane and the reticular lamina and hair bundles are stimulated by the shear in the gap. Under this condition, the external drag disappears because there is no relative motion between the bulk fluid and the hair bundle. The energy loss in that system is due to the shear between the two places and the internal drag of the hair bundle is negligible. This analysis is indeed consistent with earlier reports \cite{odi2003a,NamFettiplace2008,ProdanovicGracewskiNam2015}.

\begin{figure}[h]
\begin{center}
 
\begin{minipage}[b]{7cm} \textbf{A}  \end{minipage}\ \ \begin{minipage}[b]{3cm}\textbf{B}  \end{minipage}  \\
\includegraphics[width=0.5\textwidth]{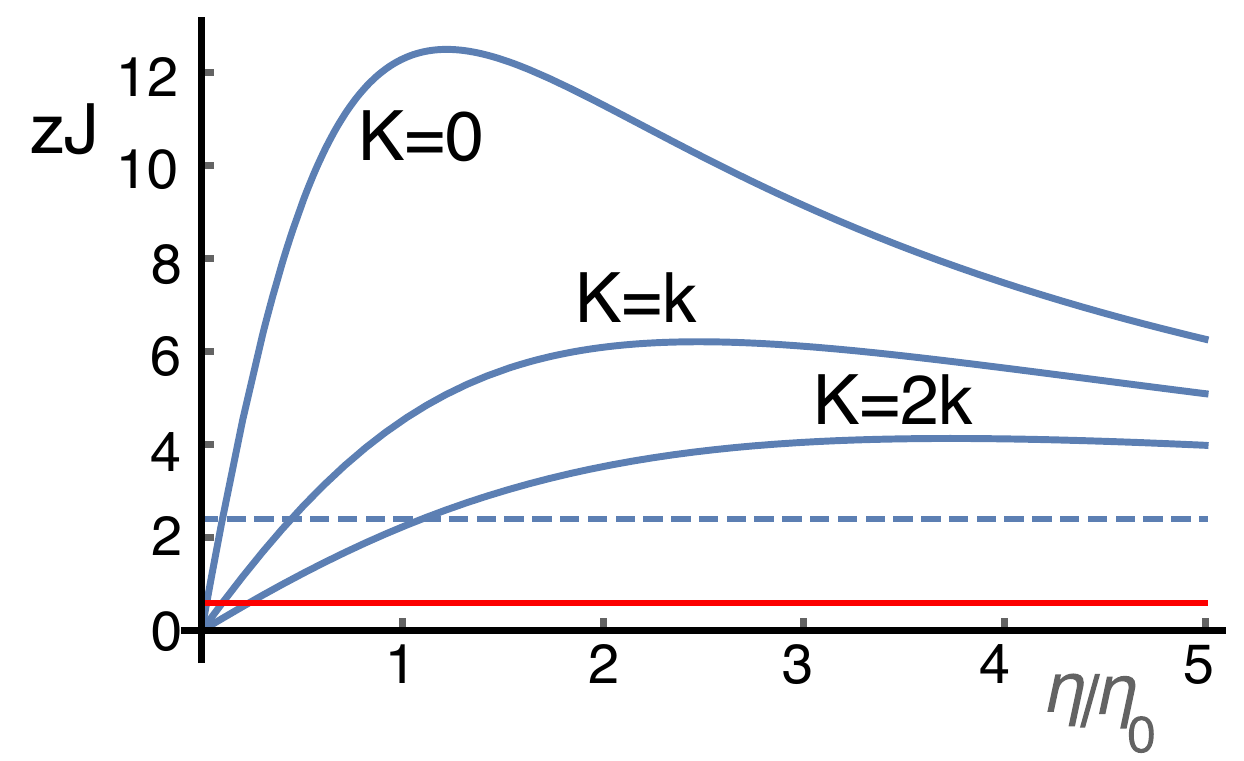}\ \ \includegraphics[width=0.45\textwidth]{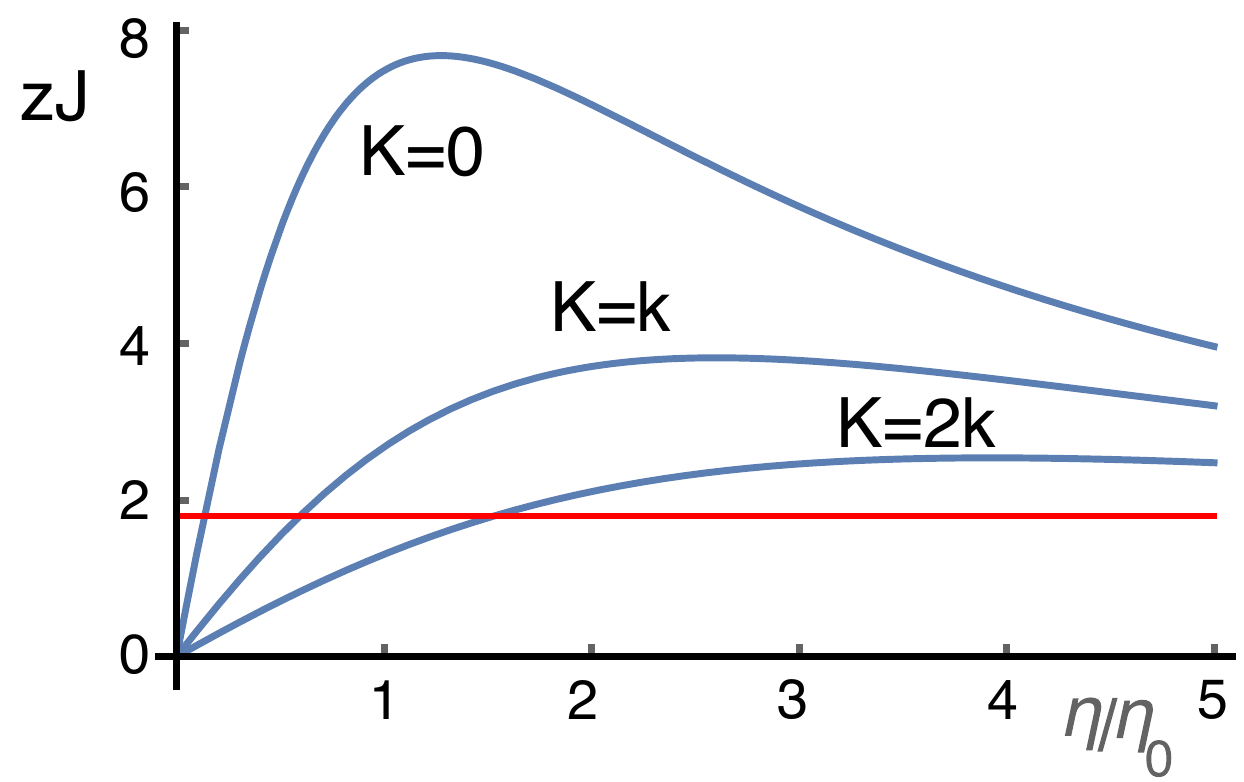}
\caption{\small{Energy balance of an OHC. \textbf{A}: Predicted at 4 kHz with the parameter values in Table \ref{table:paramvals}.  \textbf{B}: Extrapolated to 12 kHz by reducing the linear capacitance $C_0$ in half and increasing the internal stiffness $k$ twice to account for shorter cell length of basal OHCs, without accounting for the larger hair bundle conductance, which increases the efficiency. Production of dissipative energy in zJ (=$10^{-21}$J) per half cycle for 3 \% of full opening of its MET channels is plotted against drag coefficient $\eta$. The traces respectively correspond to K/k=0, 1, 2 from the top. The red full lines indicate upper bounds of energy loss due to internal drag of the hair bundle. The dashed line indicates energy loss of hair bundle with external viscous drag equivalent to a sphere of 4 $\mu$m diameter. $\eta_0=1.0\times 10^{-6}$ kg/s. 
}}
\label{fig:energy_balance}
\end{center}
\end{figure}

Assuming that the internal drag is less than 1/10 of the total, experimental data on frog saccular hair bundles \cite{Bormuth2014} can be used to estimate an upper bound of energy dissipation. Full gating of frog hair bundle stimulated at 100 $\mu$m/s with amplitude $\sim$100 nm generates viscous force $\sim$20 pN \cite{Bormuth2014}. This value is about twice as large as 80 nNs/m for frequencies higher than 2 kHz \cite{KozlovBaumgartRislerEtAl2011}. Energy dissipation based on this observed force is $10^{-19}$ J/half cycle for $f=500$ Hz for full gating. Thus 3\% gating corresponds to $3\times10^{-21}$ J/half cycle. If we extrapolate for higher frequencies, $2.4\times10^{-20}$ J/half cycle for frequency of 4 kHz. Since the drag coefficient at 4 kHz is less than that at 500 Hz \cite{KozlovBaumgartRislerEtAl2011}, this value is an overestimate. A ten-fold reduction of this value leads to $2.4$ zJ, which is given as an upper bound of energy dissipation due to internal friction of a \textit{frog} hair bundle. This value is smaller than $6$ zJ ($=6\times10^{-21}$ J) with elastic load $K=k$ (Figs.\ \ref{fig:drag_effect} C and \ref{fig:energy_balance}), and $12$ zJ without elastic load (Figs.\ \ref{fig:drag_only}B and \ref{fig:energy_balance}). 

The internal friction coefficient of a OHC hair bundle is likely much smaller than that of the frog hair bundle, given the difference in their geometries. If we assume that internal friction is due to the shear between the stereocilia, the internal friction of an OHC is $\sim$1/4 of that of a frog hair bundle (Supplement). Thus an upper bound of energy loss is $\sim 0.6$ zJ, less than 1/10 of the energy output of an OHC.   Even if we assume that the OHC is isolated and subjected to an external drag equivalent to Stokes drag for a sphere with the diameter of 4 $\mu$m, the maximal height of a hair bundle that operates at 4 kHz \cite{FurnessMahendrasingamOhashiEtAl2008} \emph{in vivo}, the comparison indicates that energy output exceeds energy loss by hair bundle drag (Fig.\ \ref{fig:energy_balance}). 

\subsubsection{Amplifier gain -- elastic energy}
The observed stiffness of the hair bundle of OHCs is between 1 to 3 mN/m~\cite{kenn-fet2005}. The hair bundle displacement required for 3 \% change in the bundle conductance (or resistance) is $\sim$0.03 nm \cite{kenn-fet2005}. Thus the energy required for changing the conductance by 3\% is $1.5$ aJ in the most efficient condition based on 1 mN/m bundle stiffness. The elastic energy output obtained from the model is $\sim$2 zJ at 4 kHz (Fig.\ \ref{fig:large_eta}B). It is 3.6 zJ even for small viscous drag (Fig.\ \ref{fig:small_eta}B). This indicates that output energy is a fraction of input energy.

Outer hair cells have, therefore, a minor effect in the elastic energy involved in the oscillation of the system. Since elastic energy is conserved it is unnecessary for the output of an amplifier to match the input.

\subsection{Energy balance in more basal cells}
The analysis described above shows that energy output exceeds energy input for an OHC that operates at 4 kHz even without mechanical resonance. More basal cells that operate at higher frequencies are more labile and harder to obtain reliable experimental data. However, the energy balance in more basal cells, which operate at higher frequencies, is of great interest because it is harder to achieve a favorable energy balance at a higher frequencies and therefore it is important for understanding the effectiveness of OHC electromotility as the basis of the cochlear amplifier \cite{mh1994,ha1992,odi2003a,Mistrik2009,maoil2013,Ramamoorthy2012a,ospeck2012}. 

If we stimulate the same cell, for example, at 12 kHz, a frequency higher by 3 times, the energy output is significantly less. However, the OHCs that operate at higher frequencies are shorter, with a lower linear membrane capacitance $C_0$ and  larger axial stiffness $k$. In addition, the hair bundle conductance is more sensitive to the strain \cite{Johnson2011}. If $C_0$ is halved and $k$ is doubled, the energy balance is still favorable (Fig.\ \ref{fig:energy_balance}) even without considering a higher sensitivity of the hair bundle transducer.  

Notice that these comparisons are made without considering mechanical resonance. The reason for excluding resonance in the present examination is the sharp sensitivity of the membrane capacitance on the resonance frequency $\omega_r$ as well as the characteristic viscoelastic frequency $\omega_\eta$ (Fig.\ \ref{fig:cnl}). 
Since OHCs in the cochlea works close to mechanical resonance frequencies, it is possible that the membrane capacitance in the operating condition is much smaller, enabling more efficient use of electrical energy \cite{mh1994}. This issue, however, cannot be addressed quantitatively without precise information, such as the values for $\omega_r$, $\omega_\eta$ as well as the sensitivity of the hair bundle.

\section{Concluding remarks}
In this paper, a simple model for electromotility of OHCs is proposed to describe its behavior in a dynamic environment. The model is consistent with the experimental data so far obtained from isolated OHCs. The model also extends the expression for the membrane capacitance, incorporating the effects of frequency, drag, elastic load, and an associated mass. Monitoring the membrane capacitance could be the easiest means of testing  the predictions. The effect of mass will be described in more detail elsewhere.

This model enables description of the receptor potential and energy production by an OHC while its hair bundle is mechanically stimulated by sinusoidal waveform. It was found that the receptor potential is more significantly affected by viscous drag than elastic load.  The model predicts that the output of elastic energy is less than the input at the hair bundle. However, the output of dissipative energy is larger than the input. Since negative drag is a usual amplifying mechanism, these results are consistent with the biological role of the OHCs as an amplifier.

\section*{Acknowledgments}
Discussion with Dr.\ Charles Steele and Yanli Wang motivated to start the present work at Stanford, partially supported by NIH 5R01DC00791008 to Stanford University. A part of the work was done during my stay at Niels Bohr Institute, Copenhagen and Max Planck Institute for Physics of Complex Systems, Dresden. It was also supported in part by the Intramural Research Program of the NIH, NIDCD. I acknowledge Drs.\ Charles Steele, Anthony Ricci, Thomas Heimburg, and Frank J\"ulicher for their interest and comments. I also thank Dr.\ Jong-Hoon Nam of Syracuse University for discussion regarding hair bundle drag and Richard Chadwick, Thomas Friedman of NIH, and three anonymous reviewers for useful comments.


\begin{thebibliography}{41}
\providecommand{\url}[1]{\texttt{#1}}
\providecommand{\urlprefix}{ }

\bibitem[Ashmore(1987)]{a1987}
Ashmore, J.~F., 1987.
\newblock A fast motile response in guinea-pig outer hair cells: the molecular
  basis of the cochlear amplifier.
\newblock \emph{J. Physiol.} 388:323--347.

\bibitem[Santos-Sacchi(1991)]{s1991}
Santos-Sacchi, J., 1991.
\newblock Reversible inhibition of voltage-dependent outer hair cell motility
  and capacitance.
\newblock \emph{J. Neurophysiol.} 11:3096--3110.

\bibitem[Dong and Olson(2013)]{Dong2013}
Dong, W., and E.~S. Olson, 2013.
\newblock Detection of cochlear amplification and its activation.
\newblock \emph{Biophys J} 105:1067--1078.

\bibitem[Zheng et~al.(2000)Zheng, Shen, He, Long, Madison, and
  Dallos]{zshlmd2000}
Zheng, J., W.~Shen, D.~Z.-Z. He, K.~B. Long, L.~D. Madison, and P.~Dallos,
  2000.
\newblock Prestin is the motor protein of cochlear outer hair cells.
\newblock \emph{Nature} 405:149--155.

\bibitem[Oliver et~al.(2001)Oliver, He, Klocker, Ludwig, Schulte, Waldegger,
  Ruppersberg, Dallos, and Fakler]{oliv-fakl2001}
Oliver, D., D.~Z. He, N.~Klocker, J.~Ludwig, U.~Schulte, S.~Waldegger, J.~P.
  Ruppersberg, P.~Dallos, and B.~Fakler, 2001.
\newblock Intracellular anions as the voltage sensor of prestin, the outer hair
  cell motor protein.
\newblock \emph{Science} 292:2340--2343.

\bibitem[Song and Santos-Sacchi(2013)]{Song2013}
Song, L., and J.~Santos-Sacchi, 2013.
\newblock Disparities in voltage-sensor charge and electromotility imply slow
  chloride-driven state transitions in the solute carrier SLC26a5.
\newblock \emph{Proc Natl Acad Sci U S A} 110:3883--3888.

\bibitem[Iwasa(1993)]{i1993}
Iwasa, K.~H., 1993.
\newblock Effect of stress on the membrane capacitance of the auditory outer
  hair cell.
\newblock \emph{Biophys. J.} 65:492--498.

\bibitem[Tolomeo and Steele(1995)]{ts1995}
Tolomeo, J.~A., and C.~R. Steele, 1995.
\newblock Orthotropic piezoelectric properties of the cochlear outer hair cell
  wall.
\newblock \emph{J. Acoust. Soc. Am.} 97:3006--3011.

\bibitem[Iwasa(2001)]{i2001}
Iwasa, K.~H., 2001.
\newblock A two-state piezoelectric model for outer hair cell motility.
\newblock \emph{Biophys. J.} 81:2495--2506.

\bibitem[Hallworth(1995)]{hal1995}
Hallworth, R., 1995.
\newblock Passive compliance and active force generation in the guinea pig
  outer hair cell.
\newblock \emph{J. Neurophysiol.} 74:2319--2328.

\bibitem[Iwasa and Adachi(1997)]{ia1997}
Iwasa, K.~H., and M.~Adachi, 1997.
\newblock Force generation in the outer hair cell of the cochlea.
\newblock \emph{Biophys. J.} 73:546--555.

\bibitem[Dallos et~al.(1993)Dallos, Hallworth, and Evans]{dhe1993}
Dallos, P., R.~Hallworth, and B.~N. Evans, 1993.
\newblock Theory of electrically driven shape changes of cochlear outer hair
  cells.
\newblock \emph{J. Neurophysiol.} 70:299--323.

\bibitem[Rabbitt et~al.(2009)Rabbitt, Clifford, Breneman, Farrell, and
  Brownell]{Rabbitt2009}
Rabbitt, R.~D., S.~Clifford, K.~D. Breneman, B.~Farrell, and W.~E. Brownell,
  2009.
\newblock Power efficiency of outer hair cell somatic electromotility.
\newblock \emph{PLoS Comput Biol} 5:e1000444.

\bibitem[Ramamoorthy and Nuttall(2012)]{Ramamoorthy2012a}
Ramamoorthy, S., and A.~L. Nuttall, 2012.
\newblock Outer hair cell somatic electromotility in vivo and power transfer to
  the organ of Corti.
\newblock \emph{Biophys J} 102:388--398.

\bibitem[Frank et~al.(1999)Frank, Hemmert, and Gummer]{fhg1999}
Frank, G., W.~Hemmert, and A.~W. Gummer, 1999.
\newblock Limiting dynamics of high-frequency electromechanical transduction of
  outer hair cells.
\newblock \emph{Proc. Natl. Acad. Sci. USA} 96:4420--4425.

\bibitem[Housley and Ashmore(1992)]{ha1992}
Housley, G.~D., and J.~F. Ashmore, 1992.
\newblock Ionic currents of outer hair cells isolated from the guinea-pig
  cochlea.
\newblock \emph{J. Physiol.} 448:73--98.

\bibitem[Johnson et~al.(2011)Johnson, Beurg, Marcotti, and
  Fettiplace]{Johnson2011}
Johnson, S.~L., M.~Beurg, W.~Marcotti, and R.~Fettiplace, 2011.
\newblock Prestin-driven cochlear amplification is not limited by the outer
  hair cell membrane time constant.
\newblock \emph{Neuron} 70:1143--1154.

\bibitem[Ospeck and Iwasa(2012)]{ospeck2012}
Ospeck, M., and K.~H. Iwasa, 2012.
\newblock How close should the outer hair cell {RC} roll-off frequency be to
  the characteristic frequency? [Correction: 103:846-847].
\newblock \emph{Biophys. J} 102:1767--1774.

\bibitem[Adachi and Iwasa(1999)]{ai1999}
Adachi, M., and K.~H. Iwasa, 1999.
\newblock Electrically driven motor in the outer hair cell: Effect of a
  mechanical constraint.
\newblock \emph{Proc. Natl. Acad. Sci. USA} 96:7244--7249.

\bibitem[Ospeck et~al.(2003)Ospeck, Dong, and Iwasa]{odi2003a}
Ospeck, M., X.-X. Dong, and K.~H. Iwasa, 2003.
\newblock Limiting frequency of the cochlear amplifier based on electromotility
  of outer hair cells.
\newblock \emph{Biophys. J.} 84:739--749.

\bibitem[Mistr\'ik et~al.(2009)Mistr\'ik, Mullaley, Mammano, and
  Ashmore]{Mistrik2009}
Mistr\'ik, P., C.~Mullaley, F.~Mammano, and J.~Ashmore, 2009.
\newblock Three-dimensional current flow in a large-scale model of the cochlea
  and the mechanism of amplification of sound.
\newblock \emph{J R Soc Interface} 6:279--291.

\bibitem[Maoil\'eidigh and Hudspeth(2013)]{maoil2013}
Maoil\'eidigh, D.~O., and A.~J. Hudspeth, 2013.
\newblock Effects of cochlear loading on the motility of active outer hair
  cells.
\newblock \emph{Proc. Natl. Acad. Sci. USA} 110:5474--5479.

\bibitem[Cole(1968)]{cole}
Cole, K.~S., 1968.
\newblock Membranes, ions, and impulses.
\newblock University of California Press, Berkely, CA.

\bibitem[Sokabe et~al.(1991)Sokabe, Sachs, and Jing]{sokabe-sachs1991}
Sokabe, M., F.~Sachs, and Z.~Q. Jing, 1991.
\newblock Quantitative video microscopy of patch clamped membranes stress,
  strain, capacitance, and stretch channel activation.
\newblock \emph{Biophys. J.} 59:722--728.

\bibitem[Iwasa(2010)]{i2010}
Iwasa, K.~H., 2010.
\newblock Chapter 6. {E}lectromotility of outer hair cells.
\newblock \emph{In} P.~A. Fuchs, editor, The {O}xford {H}andbook of {A}uditory
  {S}cience volume 1: {T}he {E}ar, Oxford University Press, Oxford, UK,
  179--212.

\bibitem[Iwasa(1997)]{i1997}
Iwasa, K.~H., 1997.
\newblock Current noise spectrum and capacitance due to the membrane motor of
  the outer hair cell: theory.
\newblock \emph{Biophys. J.} 73:2965--2971.

\bibitem[Gale and Ashmore(1997)]{ga1997}
Gale, J.~E., and J.~F. Ashmore, 1997.
\newblock An intrinsic frequency limit to the cochlear amplifier.
\newblock \emph{Nature} 389:63--66.

\bibitem[Dong et~al.(2000)Dong, Ehrenstein, and Iwasa]{Dong2000}
Dong, X., D.~Ehrenstein, and K.~H. Iwasa, 2000.
\newblock Fluctuation of motor charge in the lateral membrane of the cochlear
  outer hair cell.
\newblock \emph{Biophys J} 79:1876--1882.

\bibitem[Salt et~al.(1989)Salt, Inamura, Thalmann, and Vora]{salt1989}
Salt, A.~N., N.~Inamura, R.~Thalmann, and A.~Vora, 1989.
\newblock Calcium gradients in inner ear endolymph.
\newblock \emph{Am. J. Otol.} 10:371--375.

\bibitem[Santos-Sacchi et~al.(1998)Santos-Sacchi, Kakehata, Kikuchi, Katori,
  and Takasaka]{skkkt1998}
Santos-Sacchi, J., S.~Kakehata, T.~Kikuchi, Y.~Katori, and T.~Takasaka, 1998.
\newblock Density of motility-related charge in the outer hair cell of the
  guinea pig is inversely related to best frequency.
\newblock \emph{Neurosci Lett.} 256:155--158.

\bibitem[Freeman and Weiss(1988)]{fw1988}
Freeman, D.~M., and T.~F. Weiss, 1988.
\newblock The role of fluid inertia in mechanical stimulation of hair cells.
\newblock \emph{Hearing Res.} 35:201--208.

\bibitem[Gold(1948)]{g1948}
Gold, T., 1948.
\newblock Hearing. {II}. The physical basis of the action of the cochlea.
\newblock \emph{Proc. Roy. Soc., B} 135:492--498.

\bibitem[Kozlov et~al.(2011)Kozlov, Baumgart, Risler, Versteegh, and
  Hudspeth]{KozlovBaumgartRislerEtAl2011}
Kozlov, A.~S., J.~Baumgart, T.~Risler, C.~P.~C. Versteegh, and A.~J. Hudspeth,
  2011.
\newblock Forces between clustered stereocilia minimize friction in the ear on
  a subnanometre scale.
\newblock \emph{Nature} 474:376--379.

\bibitem[Bormuth et~al.(2014)Bormuth, Barral, Joanny, J{\"{u}}licher, and
  Martin]{Bormuth2014}
Bormuth, V., J.~Barral, J.-F. Joanny, F.~J{\"{u}}licher, and P.~Martin, 2014.
\newblock Transduction channels' gating can control friction on vibrating
  hair-cell bundles in the ear.
\newblock \emph{Proc Natl Acad Sci U S A} 111:7185--7190.

\bibitem[Nam and Fettiplace(2008)]{NamFettiplace2008}
Nam, J.-H., and R.~Fettiplace, 2008.
\newblock Theoretical conditions for high-frequency hair bundle oscillations in
  auditory hair cells.
\newblock \emph{Biophys J} 95:4948--4962.

\bibitem[Prodanovic et~al.(2015)Prodanovic, Gracewski, and
  Nam]{ProdanovicGracewskiNam2015}
Prodanovic, S., S.~Gracewski, and J.-H. Nam, 2015.
\newblock Power dissipation in the subtectorial space of the mammalian cochlea
  is modulated by inner hair cell stereocilia.
\newblock \emph{Biophys J} 108:479--488.

\bibitem[Kennedy et~al.(2005)Kennedy, Crawford, and Fettiplace]{kenn-fet2005}
Kennedy, H.~J., A.~C. Crawford, and R.~Fettiplace, 2005.
\newblock Force generation by mammalian hair bundles supports a role in
  cochlear amplification.
\newblock \emph{Nature} 433:880--883.

\bibitem[Mountain and Hubbard(1994)]{mh1994}
Mountain, D.~C., and A.~E. Hubbard, 1994.
\newblock A piezoelectric model of outer hair cell function.
\newblock \emph{J. Acoust. Soc. Am.} 95:350--354.

\bibitem[M{\"{u}}ller(1991)]{Mueller1991a}
M{\"{u}}ller, M., 1991.
\newblock Frequency representation in the rat cochlea.
\newblock \emph{Hear Res} 51:247--254.

\bibitem[Furness et~al.(2008)Furness, Mahendrasingam, Ohashi, Fettiplace, and
  Hackney]{FurnessMahendrasingamOhashiEtAl2008}
Furness, D.~N., S.~Mahendrasingam, M.~Ohashi, R.~Fettiplace, and C.~M. Hackney,
  2008.
\newblock The dimensions and composition of stereociliary rootlets in mammalian
  cochlear hair cells: comparison between high- and low-frequency cells and
  evidence for a connection to the lateral membrane.
\newblock \emph{J Neurosci} 28:6342--6353.

\bibitem[Lim(1986)]{lim1986}
Lim, D.~J., 1986.
\newblock Functional structure of the organ of Corti: a review.
\newblock \emph{Hear Res} 22:117--146.

\end{thebibliography}

\renewcommand{\thefigure}{A\arabic{figure}}
\setcounter{figure}{0}
\section*{Supplement: Internal drag of hair bundles}
Here internal drag of a hair bundle is estimated assuming it is due to the shear between stereocilia in a hair bundle.

\subsection*{Shear between stereocilia}
To provide an estimate, the shear between two stereocilia is approximated by the shear between two plates, assuming the separation is the same as the rootlet separation $s$, and the widths of the plates is given by the diameter $d$ of the stereocilia, ignoring end effects. This model overestimates the separation because the nearest distance between the stereocilia is less than $s$, resulting in an underestimate of the drag. At the same time the planar model overestimates the drag because end effects are not considered.

If the angle $\theta$ of bending and the separation $s$ between the stereocilia are small, viscous drag $F_p$ between a pair of stereocilia can be expressed,
\begin{eqnarray*}
F_p/v_s\approx \eta (hd) /s,
\end{eqnarray*}
where $v_s$ is the speed of the relative motion of the stereocilia, $s$ root separation, $h$ height of the gap,  $d$ stereocila diameter, and $h_0$ tip height (Fig.\ A1). Since $v_s\approx s \cdot d\theta/dt$ and the tip velocity is expressed by $v_t\approx h_0 \cdot d\theta/dt$, the drag coefficient  $F_p/v_t$ with respect to tip velocity does not depend on the separation $s$. If the bundle has $N$ sliding pairs, the drag coefficient $F_b/v_t$ of the bundle is,
\begin{equation*}
F_b/v_t\approx N\eta d \cdot (h/h_0).
\end{equation*}

\begin{figure}
\begin{center}
\begin{tikzpicture}
\draw[ultra thick] (-2.5,0) -- (2.5,0);
\draw[ultra thick](-0.6,0)--(-0.2,2);
\node at (-0.1,2.3) {$h$};
\draw[ultra thin](-0.6,0)--(-0.6,3);
\draw (-0.3,1.2) arc (70:90:1);
\node at (-0.8,1) {$\theta$};
\draw[ultra thick](0.6,0)--(1.4,4);
\node at (1.8,3.8) {$h_0$};
\draw[thick,->](0.5,4)--(1.3,4);
\node at (0.8,3.7) {$v_t$};
\draw[ultra thin] (-0.2,2)--(1.0,2);
\draw[ultra thin] (-0.2,1.95)--(0.9,1.7);
 \node at (0.3,1.6) {$\theta$};
\draw (0.5,2) arc (0:-28:0.5);
\node at (0,-0.2) {$s$};
\draw [ultra thin,->] (0.2,-0.2)--(0.6,-0.2);
\draw [ultra thin,<-] (-0.6,-0.2)--(-0.2,-0.2);
\end{tikzpicture}
\end{center}
\caption{Shear between two stereocilia}
\end{figure}
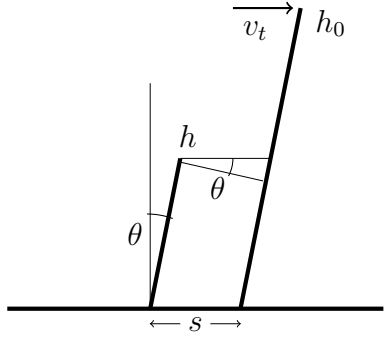

\subsection*{Frog saccular hair bundle}
Frog saccular hair bundle is modeled by a hexagonally packed hexagonal array that consists of 62 stereocilia \cite{KozlovBaumgartRislerEtAl2011}.   Let a displacement applied through the kinocilium at a vertex toward the center of the bundle.  The number of sliding pairs in this direction is 53. The number of pairs away from this direction by 60 degree is 106. Those pairs are subjected a half as large shear as those in the direction of stimulation. Thus the effective number $N$ of sliding pairs is 106. The mean height $h$ of these pairs is $\sim 0.75 h_0$. Hence
$N\cdot (h/h_0)\approx 75$.

If we choose a value 0.4 $\mu$m for the radius $d$ of stereocilia, we obtain $F_b/v_t=30 \times 10^{-9}$ Ns/m, which is about 1/3 of the total drag coefficient \cite{KozlovBaumgartRislerEtAl2011},   consistent with   the significance of end effects.  If we choose d$\le$0.1 $\mu$m, $F_b/v_t\le 7.5 \times 10^{-9}$ Ns/m, $\le$1/10 of the total drag coefficient. It is compatible with the calculation of Kozlov et al. \cite{KozlovBaumgartRislerEtAl2011}

\subsection*{OHC hair bundle}

In the cochlea of rats, 4 kHz location is at 80\% from the base \cite{Mueller1991a}, which is in the apical turn.  The ratio $h/h_0$ is typically 0.5 \cite{FurnessMahendrasingamOhashiEtAl2008}. If we assume the total number of stereocilia is about 60, similar to chinchilla hair bundles \cite{lim1986}, the number of sliding pairs is 40. If we can use d$\le$0.1$\mu$m for the effective radius of stereocilia, we obtain $F_b/v_t\le 2 \times 10^{-9}$ Ns/m.

\end{document}